\documentclass[10pt]{article}
\usepackage{latexsym}
\usepackage{amsmath,amssymb,amsfonts}
\usepackage{epsf,epsfig,graphics,graphicx}
\usepackage{color,bm}
\usepackage{dcolumn}     
\usepackage{dsfont,slashed}
\usepackage{placeins}
\usepackage{wasysym}

\newcommand{\be}{\begin{equation}}
\newcommand{\ee}{\end{equation}}
\newcommand{\bea}{\begin{eqnarray}}
\newcommand{\eea}{\end{eqnarray}}
\newcommand{\<}{\langle}
\renewcommand{\>}{\rangle}

\newcommand{\mc}{\mathcal}

\newcommand{\qvec}{\vec{q}}

\newcommand{\idnty}{\hbox{1$\!\!$1}}
\newcommand{\C}{\mathcal{C}}

\renewcommand{\l}{\left}
\renewcommand{\r}{\right}

\def\gm{\gamma}
\def\nn{\nonumber\\}

\def\eq#1{Eq.~(\ref{#1})}
\def\fig#1{Fig. \ref{#1}}
\def\tbl#1{Table \ref{#1}}
\def\sec#1{Section \ref{#1}}

\def\cyp{a}
\def\cyi{b}

\def\nic{d}
\def\hisk{e}

\newlength{\figsize}
\begin{document}
\begin{titlepage}
  \begin{center}
    \begin{LARGE}
      \textbf{Neutron electric dipole moment using $N_f{=}2{+}1{+}1$ twisted mass fermions} \\
    \end{LARGE}
  \end{center}

\vspace{.5cm}

\vspace{-0.8cm}
  \baselineskip 20pt plus 2pt minus 2pt
  \begin{center}
    \textbf{
      C.~Alexandrou$^{(\cyp, \cyi)}$,
      A.~Athenodorou$^{(\cyp, \cyi)}$,
      M.~Constantinou$^{(\cyp, \cyi)}$,
      K.~Hadjiyiannakou$^{(\cyp, \cyi)}$,
      K.~Jansen$^{(\nic)}$,
      G.~Koutsou$^{(\cyi)}$
      K.~Ottnad$^{(\cyp, \hisk)}$
      M.~Petschlies$^{(\cyi, \hisk)}$
}
  \end{center}

  \begin{center}
    \begin{footnotesize}
      \noindent 	
 	$^{(\cyp)}$ Department of Physics, University of Cyprus, P.O. Box 20537,
 	1678 Nicosia, Cyprus\\	
 	$^{(\cyi)}$ Computation-based Science and Technology Research Center, The Cyprus Institute, 20 Kavafi Str., Nicosia 2121, Cyprus \\
      $^{(\nic)}$ NIC, DESY, Platanenallee 6, D-15738 Zeuthen, Germany\\
      $^{(\hisk)}$ Helmholtz-Institut f\"ur Strahlen- und Kernphysik (Theorie) and Bethe Center for Theoretical Physics, Universit\"at Bonn, 53115 Bonn, Germany \\
     \vspace{0.2cm}
    \end{footnotesize}
  \end{center}

\begin{abstract}
We evaluate the neutron electric dipole moment  $\vert
\vec{d}_N\vert$ using lattice QCD techniques. The gauge
configurations analyzed are produced by the European Twisted Mass
Collaboration using $N_f{=}2{+}1{+}1$ twisted mass fermions at one value of
the lattice spacing of $a \simeq 0.082 \ {\rm fm}$ and a light quark
mass corresponding to $m_{\pi} \simeq 373 \ {\rm MeV}$. Our approach
to extract the neutron electric dipole moment is based on the
calculation of the $CP$-odd electromagnetic form factor $F_3(Q^2)$
for small values of the vacuum angle $\theta$ in the limit of zero
Euclidean momentum transfer $Q^2$. The limit $Q^2 \to 0$ is
realized either by adopting a parameterization of the momentum
dependence of $F_3(Q^2)$ and performing a fit, or by employing  new position space methods,
which involve the elimination of the kinematical momentum
factor in front of $F_3(Q^2)$. 
The computation in the presence of a $CP$-violating term requires the
evaluation of the  topological charge ${\cal Q}$.
This is computed by applying the cooling technique and
the gradient flow with three different actions, namely the Wilson, the Symanzik tree-level
improved and the Iwasaki action. We demonstrate that cooling and
gradient flow give equivalent results for the neutron electric dipole moment.
Our analysis  yields a value of $\vert \vec{d}_N\vert=0.045(6)(1)
\ \bar{\theta} \ e \cdot {\rm fm}$ for the ensemble with $m_\pi=373$~MeV considered. 
 \begin{center}
  \today
 \end{center}
\end{abstract}

\end{titlepage}

\tableofcontents

\newpage

\section{Introduction}
\label{sec:Introduction}
 The discrete symmetries of parity $P$, charge conjugation
$C$ and time-reversal $T$ play an important role in the allowed
phenomenology described by the Standard Model (SM) of particle physics.
 An experimental observation of a non-vanishing electric-dipole moment 
on the neutron would directly signal  violation of both $P$ and  $T$ symmetries.
 Violations of $P$ and $T$ can occur in both strong and electroweak 
sectors of the Standard Model. \par

So far, a non-vanishing neutron electric dipole moment (nEDM) has not been
reported and current bounds are still several orders of magnitude
above what one expects from $CP$ violation induced by weak
interactions~{\cite{Dar:2000tn}}, making, thus, nEDM investigations an interesting probe for  Beyond the Standard Model (BSM) physics~\cite{Pospelov:2005pr}. Several experiments are
under way to improve the  upper bound on the
nEDM, $\vec{d}_N$, with the best experimental upper limit
being~\cite{Helaine:2014ona,Baker:2006ts,Baker:2007df} 
\be
\vert \vec{d}_N \vert  < 2.9 \times 10^{-13} e \cdot {\rm fm} \ (90\% \ {\rm CL})\,.
\label{eq:nEDM_smaller_experimental_upperbound}
\ee 
This result has been extracted at the Institut Laue-Langevin (ILL) reactor in Grenoble by
storing ``ultra-cold'' neutrons and measuring the change in the
neutron spin precession frequency in a weak magnetic field when a
strong, parallel background electric field is reversing its own sign. \par

To examine theoretically how an nEDM may arise, we start with the
$CP$-conserving QCD Lagrangian density, which in Euclidean space is
given by
\bea
  {\cal L}_{\rm QCD} \left( x \right)= \frac{1}{2 g^2} 
{\rm Tr} \left[ G_{\mu \nu} \left( x \right) G_{\mu \nu} \left( x \right) \right] + 
\sum_{f} {\overline \psi}_{f} \left( x \right) (\gamma_{\mu} D_{\mu} + m_f) \psi_{f}\left( x \right)\,,
\label{eq:QCD_Lagrangian}
\eea
where $\psi_f$ denotes a fermion field of flavor $f$ with bare mass
$m_f$ and $G_{\mu \nu}$ is the gluon field tensor. 
\eq{eq:QCD_Lagrangian} is invariant under $P$ and $T$ transformations
and, thus, cannot lead to  a non-vanishing nEDM. The QCD Lagrangian
can be generalized by including an additional $CP$-violating
interaction (Chern-Simons) term given by 
\bea
{\cal L}_{\rm CS} \left( x \right) \equiv - i \theta q \left( x \right)\,.
\label{eq:QCD_Chern_Simons}
\eea
The  so called $\theta$-parameter controls the strength of the
$CP$-breaking and $q \left( x \right)$ is the topological charge
density, which in Euclidean space is defined as 
\bea
  q \left( x \right) = \frac{1}{ 32 \pi^2} \epsilon_{\mu \nu \rho \sigma} 
{\rm Tr} \left[ G_{\mu \nu} \left( x \right) G_{\rho \sigma} \left( x \right) \right]\,,
\label{eq:Topological_Charge_Density}
\eea
where $\epsilon_{\mu \nu \rho \sigma}$ is the totally antisymmetric
tensor. Although the $CP$-violating term in \eq{eq:QCD_Chern_Simons}
does not modify the equations of motion since it can be expressed as a
total divergence, it has observable consequences. In particular,
it leads to a non-zero value for the  nEDM. In this work, we consider a quantum
field theory described by the $C$-even Lagrangian density
\bea
{\cal L} \left( x \right) = {\cal L}_{\rm QCD} \left( x \right) + {\cal L}_{\rm CS}\left( x \right)\,,
\label{eq:Lagrangian_Density}
\eea
with $\theta$  taken as a small continuous parameter enabling us
to perturbatively expand in terms of $\theta$ and only keep first
order contributions. 

$CP$ violation in the electroweak sector is  observed in K and B meson
decays and is accounted for by the phase of the CKM matrix. However,
this $CP$-violating phase alone cannot explain the baryon asymmetry of
the universe  suggesting that there maybe additional sources of $CP$
violation. If one considers the electroweak sector of the Standard
Model, the Lagrangian in \eq{eq:Lagrangian_Density} gets a contribution from the quark mass matrix $M$,
arising from Yukawa couplings to the Higgs field
\bea
    \bar\psi^R_{f_1}(x) M_{{f_1}{f_2}}\psi^L_{f_2}(x) + \bar\psi^L_{f_1}(x)
M^{\dag}_{{f_1}{f_2}}\psi^R_{f_2}(x)\,,
\label{eq:Yukawa_Couplings}
\eea
with $\psi^L_f$ and $\psi^R_f$ being the left and right handed quark
fields with flavour indices $f$. If one performs a  $U(1)_A$ chiral
transformation an additional $\epsilon_{\mu \nu \rho \sigma} {\rm Tr}
\left[ G_{\mu \nu} G_{\rho \sigma} \right]$ contribution is introduced
because of the chiral anomaly. Hence, the parameter $\theta$ shifts to
${\overline \theta} = \theta + \arg\det M$ where now ${\overline \theta}$
describes the $CP$-violating parameter of the extended strong and
electroweak symmetry. In addition to the experimental investigations
that give an upper bound in the nEDM, several model studies
\cite{Baluni:1978rf,Crewther:1979pi,Shifman:1979if,Schnitzer:1983pb,Shabalin:1982sg,Musakhanov:1984qy,Cea:1984qv,Morgan:1986yy,McGovern:1992bk,DiVecchia:1980gi},
as well as more recent effective field theory calculations
\cite{Pich:1991fq,Borasoy:2000pq,Hockings:2005cn,Narison:2008jp,Ottnad:2009jw,deVries:2010ah,Mereghetti:2010kp,deVries:2012ab,Guo:2012vf,Akan:2014yha},
have attempted to provide a value for the nEDM. They report values in the range of
\bea
\vert { \vec{d}_N} \vert \sim {\bar{\theta}} \cdot {\cal O} \left( 10^{-2} - 10^{-4} \right) e \cdot {\rm fm}\,.
\label{eq:Model_Dipole_Moment}
\eea
Using the experimental upper bound
(\eq{eq:nEDM_smaller_experimental_upperbound}) and the above prediction
we obtain a bound of the order 
${\bar \theta} \lesssim {\cal O} \left( 10^{-9} - 10^{-11} \right)$.
Hence, according to these models, ${\bar \theta}$ is
indeed very small and possibly zero with the latter resulting in a vanishing value of the nEDM.

Therefore, either $\theta$ and $\arg\det M$ are
small or they cancel each other at a level  that the experimental upper
bound on the nEDM is satisfied. No matter which one of these two cases
holds, one needs to be able to explain why nature chooses such a small
value for ${\overline \theta}$. This is what is referred to as the ``strong
$CP$ problem''. Attempts to  explain the smallness of the nEDM invoke
new physics as for instance  the Peccei-Quinn
mechanism~\cite{Peccei:1977hh,Peccei:1988ci}, which requires the existence
of the axion that to date has not been  observed. For the purposes of this work
we assume that $\theta$ is small and keep only leading order
contributions in what follows.

The effective Lagrangian giving rise to the nEDM at leading order in ${\bar{\theta}}$ can be written as ~\cite{Pospelov:2005pr,Jarlskog:1988}
\bea
- {\bar{\theta}} \frac{F_3(Q^2)}{4m_N}\bar{u}_N(p_f)\sigma^{\mu \nu} \gamma_5 u_N(p_i)F^{\mu \nu} \, ,
\eea
in Euclidean space.
We denote by ${u_N}(p)$ the nucleon spinor and  by $F_{\mu \nu}$ the
electromagnetic field tensor, while
 $\sigma_{\mu\nu}=[\gm_{\mu},\gm_{\nu}]/2$
and $p_f (p_i)$ is the final (initial) momentum.  $m_N$ denotes the mass of the neutron, $Q^2{=}-q^2$ the four-momentum 
transfer in Euclidean space ($q{=}p_f-p_i$) and
$F_3(Q^2)$ is the $CP$-odd form factor.
The  nEDM,
$\vec{d}_N$ is then given by~\cite{Pospelov:2005pr,Jarlskog:1988}
\bea
\vert \vec{d}_N \vert = {\bar{\theta}} \lim_{Q^2 \to 0} \frac{\vert F_3(Q^2) \vert}{2 m_N}\,,
\label{eq:dN}
\eea
to leading order in ${\bar{\theta}}$. 
 In a theory with $CP$
violation we can, therefore, calculate the electric dipole
moment by evaluating the zero momentum transfer limit of the
$CP$-odd form factor. This provides the framework on which our
investigation will be based. As will be explained in Section~\ref{sec:Plateau_method_to_extract_F_3}, the $CP$-violating nucleon matrix element,
decomposes to $Q_k F_3(Q^2)$ ($k{=}1,2,3$) and not
$F_3(Q^2)$ alone, hindering a direct extraction of $F_3(0)$. We adopt
two approaches to determine $F_3(0)$: The first one that is commonly
applied in lattice QCD computations of form factors, is to take a
suitable parameterization of the $Q^2$-dependence. We take a dipole
form for the $Q^2$-dependence of $F_3(Q^2)$ and perform a fit to extract its value at $Q^2= 0$. The second
approach is a new method that we have recently developed to compute form factors directly at
$Q^2=0$,
extracted from matrix elements that involve a multiplicative
kinematical factor of $\vec{Q}$~\cite{Alexandrou:2014exa}. To this end we use two techniques,
the so-called {``application of the derivative to the ratio''} as well as the {``elimination of the momentum in the plateau region''} both yielding $F_3(0) $ without any model assumption  on its $Q^2$-dependence. \par  

\begin{figure}[htb]
\vspace{-1cm}
\centerline{\hspace{0.0cm}\includegraphics[scale=0.45,angle=270]{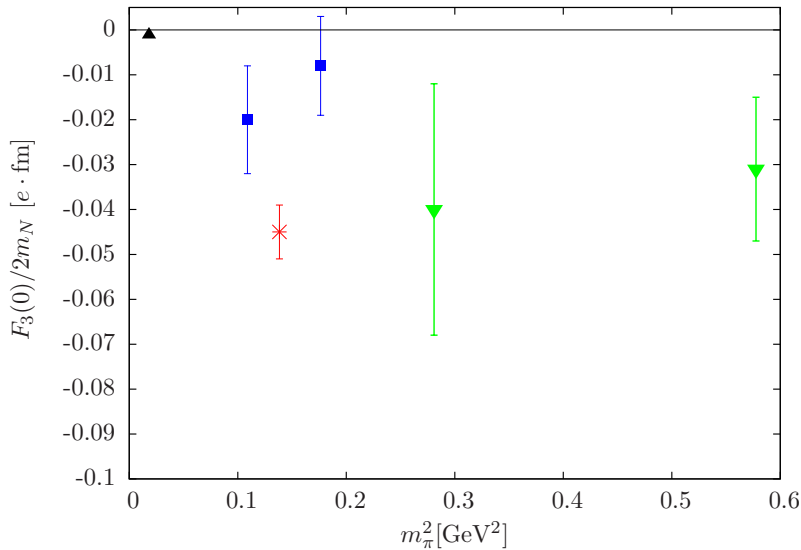}}
\vspace{-0.5cm}
\caption{\label{fig:Lattice_Comparison} $F_3(0)/(2 m_N)$
versus the pion mass squared ($m_\pi^2$). Results using $N_f{=}2{+}1{+}1$
twisted mass fermions (this work) are shown with a red asterisk. We also
show  results for $N_f{=}2{+}1$
domain wall fermions~\cite{Shintani:2014zra} at {$a \simeq 0.11$fm}
where  the $CP$-odd $F_3(Q^2)$ was evaluated and
$F_3(0)$ was determined by fitting its $Q^2$-dependence  (blue squares). Results obtained with $N_f{=}2$ Clover fermions at $a \simeq 0.11$fm using a
background electric field method are shown with
downward green triangles~\cite{Shintani:2008nt}. All errors shown are
statistical. A value determined in chiral perturbation theory at next to leading order is
shown with the black  triangle~\cite{Ottnad:2009jw}.}
\end{figure}

In order to compute the $CP$-violating matrix element and extract the
$CP$-odd form factor $F_3(Q^2)$ one needs the evaluation of the topological
charge, ${\cal Q}$. In this work we employ the field theoretic definition and use cooling and the gradient flow to smooth the gauge
links and obtain a well-defined, renormalization-free topological
charge~\cite{Luscher:2010iy}. We consider the Wilson, the Symanzik tree-level improved and
the Iwasaki actions for the cooling and the gradient flow. {Smoothing with different actions leads to observables, such as the nEDM, with potentially different lattice artifacts due to the fact that the topological charge between different actions differs only by lattice artifacts \cite{Alexandrou:2015yba}}. {The results on the nEDM arising from these different definitions of the topological
charge are compared and found compatible, demonstrating that lattice artifacts in the definition of the topological charge are small}. In \fig{fig:Lattice_Comparison} 
we show our final result for $F_3(0)/(2 m_N)$ as well as
results from other lattice investigations with dynamical quarks that
have been obtained using $\theta$ as a real parameter in the QCD
Lagrangian {keeping the comparison within a similar lattice methodology where lattice systematics are expected to be similar}. We note that results obtained using
formulations with an imaginary $\theta$ such as those by Guo 
{\it et al.}~\cite{Guo:2015tla} 
 are in agreement with our value. However, in
Fig.~\ref{fig:Lattice_Comparison} and in what follows we choose to compare results using a setup with  a
real value of $\theta$. We also  do not show
results obtained in  the quenched
approximation~\cite{Shindler:2015aqa}.  Our value displayed in Fig.~\ref{fig:Lattice_Comparison} is the
weighted average of the values obtained using different methods for
extracting $F_3(0)$ (see Section~\ref{sec:Results on nEDM}), while
for
the topological charge, we employ the Iwasaki action for its
definition, which is the same as the gauge action used in the
simulations. For the data shown in \fig{fig:Lattice_Comparison}
we use gradient flow for the computation of the topological charge,
which are, however, equivalent to the ones using cooling results (see Section
\ref{sec:The_Topological_Charge}). We also include, for comparison,  the
value of the nEDM arising from a recent chiral perturbation theory
analysis {at next to leading order}~\cite{Ottnad:2009jw}.  \par  

The article is organized as follows: In \sec{sec:Simulation_Details}
we discuss the lattice formulation used for the production of
configurations, as well as, the parameters of the ensemble
analyzed. Subsequently, in \sec{sec:nEDM_matrix_elements}, we give the
decomposition of the nucleon matrix elements in the
presence of a $CP$-violating term in the Lagrangian. In 
\sec{sec:correlators}, \sec{sec:Plateau_method_to_extract_alpha} and
\sec{sec:Plateau_method_to_extract_F_3} we explain the computation of
correlation functions and the extraction  of the form factor
$F_3(Q^2)$ in lattice QCD, including the techniques used to obtain
$F_3(0)$. In \sec{sec:The_Topological_Charge}, we discuss the
computation of the topological charge using both the cooling
and the gradient flow methods. Finally, in Section~\ref{sec:Results on
  nEDM} we present our results for the nEDM and in
Section~\ref{sec:Conclusions}  we provide our conclusions. \par 

\section{Simulation}
\label{sec:Simulation_Details}
\subsection{Formulation}
\label{sec:formulation}
We discuss here the formulation  used for the production of the
gauge configurations analyzed in this work. For the gluonic action we use
the Iwasaki improved action given by~\cite{Iwasaki:1996sn}
\bea
    S_G =  \frac{\beta}{3}\sum_x\Biggl(  c_0\sum_{\substack{
      \mu,\nu=1 \\ \mu<\nu}}^4\left \{1-{\rm Re } {\rm Tr} (U^{1\times1}_{x,\mu,\nu})\right \}\Bigr. 
     \Bigl.+
    c_1\sum_{\substack{\mu,\nu=1\\\mu\neq\nu}}^4\left \{1
    -{\rm Re } {\rm Tr}(U^{1\times2}_{x,\mu,\nu})\right \}\Biggr)\,,
  \label{eq:Sg_Iwasaki}  
\eea
where $\beta=6/g^2$, $U^{1\times1}_{x,\mu,\nu}$ is the plaquette and
$U^{1\times2}_{x,\mu,\nu}$ rectangular $(1\times2)$ Wilson loops. The
Symanzik coefficients are set to $c_0=3.648$ and $c_{1}=-0.331$ and
obey the normalization $c_0+8 c_1=1$ ensuring that the Iwasaki action
tends to the right Yang-Mills action in the continuum limit. 
For the discretization of the
fermionic action we consider the twisted mass formulation of lattice
QCD~\cite{Frezzotti:2000nk,Frezzotti:2003ni}. Although we are extracting $P$-odd quantities, all  expectation values involve $P$-even operators and, thus, this formulation provides automatic ${\it O}(a)$ improvement at maximal twist~\cite{Frezzotti:2005gi}. In addition it provides infrared regularization of small eigenvalues and allows for efficient simulations with dynamical
fermions. For the mass-degenerate doublet of light quarks we use the
action 
\be
S_F^{(l)}\left[\chi^{(l)},\overline{\chi}^{(l)},U \right]= a^4\sum_x  \overline{\chi}^{(l)}(x)\bigl(D_W[U] + m_{0,l} + i \mu_l \gamma_5\tau^3  \bigr ) \chi^{(l)}(x)\,,
\label{eq:Action_TML}
\ee
where $\tau^3$ is the third Pauli matrix acting in the flavour space,
$m_{0,l}$ the bare untwisted light quark mass and $\mu_l$ the bare
twisted light quark mass.  The massless Wilson-Dirac operator is given
by
\be \label{eq:wilson_term}
D_W[U] = \frac{1}{2} \gamma_{\mu}(\vec{\nabla}_{\mu} + \vec{\nabla}_{\mu}^{*})
-\frac{ar}{2} \vec{\nabla}_{\mu} \vec{\nabla}^*_{\mu}\,,
\ee
with
\be
\vec{\nabla}_\mu \psi(x)= \frac{1}{a}\biggl[U_\mu(x)\psi(x+a\hat{\mu})-\psi(x)\biggr]
\hspace*{0.5cm} {\rm and}\hspace*{0.5cm} 
\vec{\nabla}^*_{\mu}\psi(x)=-\frac{1}{a}\biggl[U^\dagger_{\mu}(x-a\hat{\mu})\psi(x-a\hat{\mu})-\psi(x)\biggr]\,,
\ee
the forward and backward covariant derivatives, respectively. The
action is written in terms of the fields in the ``twisted basis'',
$\chi^{(l)}$, which are related to the fields in the physical basis,
$\psi^{(l)}$, at maximal twist through the transformations
\be
\psi^{(l)}(x)=\frac{1}{\sqrt{2}}\left(\idnty+ i \tau^3\gamma_5\right) \chi^{(l)}(x),\qquad {\rm and} \qquad
\overline{\psi}^{(l)}(x)=\overline{\chi}^{(l)}(x) \frac{1}{\sqrt{2}}\left(\idnty + i \tau^3\gamma_5\right)\,.
\ee
 Apart from the doublet of light quarks, we also include a twisted
 heavy mass-split doublet $\chi^{(h)} = \left(\chi_c,\chi_s \right)$
 for the strange and charm quarks~\cite{Frezzotti:2003xj}. The associated action is 
\be
S_F^{(h)}\left[\chi^{(h)},\overline{\chi}^{(h)},U \right]= a^4\sum_x  \overline{\chi}^{(h)}(x)\bigl(D_W[U] + m_{0,h} + i\mu_\sigma \gamma_5\tau^1 + \tau^3\mu_\delta  \bigr ) \chi^{(h)}(x)\,,
\label{eq:Action_heavy}
\ee
with $m_{0,h}$ the bare untwisted quark mass for the heavy doublet,
$\mu_\sigma$ the bare twisted mass along the $\tau^1$ direction and
$\mu_\delta$ the mass splitting in the $\tau^3$ direction. The heavy
quark fields in the twisted basis are related to those in the physical
basis at maximal twist through 
\be
\psi^{(h)}(x)=\frac{1}{\sqrt{2}}\left(\idnty+ i \tau^1\gamma_5\right) \chi^{(h)}(x),\qquad {\rm and} \qquad
\overline{\psi}^{(h)}(x)=\overline{\chi}^{(h)}(x) \frac{1}{\sqrt{2}}\left(\idnty + i \tau^1\gamma_5\right)\,.
\ee 
Unless otherwise stated, in what follows
we used the quark fields in the physical basis. The fermionic action in
\eq{eq:Action_TML} breaks parity and isospin at non-vanishing lattice
spacing with the latter inducing a cut-off effect of $ {\it{O}}(a^2)$~\cite{Frezzotti:2003ni}.

The reader can find more details on the twisted mass fermion action in
Ref.~\cite{Baron:2010bv}. Simulating a charm quark may give rise to
concerns regarding cut-off effects. The observables in this work
cannot be used to check for finite lattice spacing effects induced by
heavy sea quarks. However, analyses in
Refs.~\cite{Athenodorou:2011zp,Bruno:2014ufa,Alexandrou:2014sha} show that
such cut-off effects are small.\par 

\subsection{Simulation Details}
\label{sec:simulation_details}
A number of new techniques are implemented for the extraction of the
nEDM using gauge configurations produced with $N_f{=}2{+}1{+}1$ twisted mass
fermions~\cite{Baron:2011sf}. To explore these techniques we analyze
a single ensemble for which a large number of gauge
configurations are available allowing us to reach the required
accuracy to reliably benchmark the various methods. 
Although this is a calculation using a single ensemble, previous studies have shown that finite lattice spacing effects on e.g. the nucleon mass for $a<0.1$ fm are
smaller than our statistical errors of about $\sim 3 \%$ and we expect this to hold also here. The ensemble is the so called B55.32 in the notation of
Ref.~\cite{Baron:2011sf}, which has a lattice spacing of
$a\simeq0.082$~fm determined from the nucleon mass, pion mass 373~MeV,
and a spatial lattice extent of $L/a=32$. The parameters of the
ensemble are given in \tbl{Table:params}.\par
\begin{table}[ht]
\begin{center}
\begin{tabular}{c|llll}
\hline\hline
\multicolumn{3}{c}{$\beta=1.95$, $a=0.0823(10)$~fm,   ${r_0/a}=5.710(41)$}\\\hline
$32^3\times 64$, $L=2.6$~fm  &$a\mu$ & 0.0055  \\
                               & No. of confs & 4623  \\
                               & $a\,m_\pi$& 0.15518(21)(33)\\
                               & $Lm_\pi $    & 4.97   \\
                               & $m_\pi$ & 0.373 GeV   \\
                               & $am_N$ & 0.5072(17)   \\
                               & $m_N$ & 1.220(5) GeV \\\hline \hline
\end{tabular}
\caption{Input parameters ($\beta,L,a\mu$) of our lattice
calculation (B55.32 ensemble) including the lattice spacing, $a$,
determined from the nucleon mass. Both the neutron ($m_N$) and pion
($m_\pi$) mass are given in lattice and physical units.}
\label{Table:params}
\end{center}
\end{table}
\section{Nucleon matrix elements in the presence of the $\theta$-term}
\label{sec:nEDM_matrix_elements}
A precise determination of the neutron electric dipole moment from first principles may provide
a valuable input for future experiments seeking to observe a
non-vanishing nEDM. It has been recognized since many years that
lattice QCD provides an ideal framework for a non-perturbative
investigation of the nEDM with first attempts to evaluate it dating
back nearly three decades~\cite{Aoki:1989rx}. This first pioneering work
was based on the introduction of an external electric field and the
measurement of the associated energy shift. Although for the following
ten years there was not much progress, during the last decade the
study of nEDM has been
revived~\cite{Guadagnoli:2002nm,Facciolia:2004jz,Shintani:2005xg,Shintani:2006xr,Alles:2005ys,Shintani:2008nt,Aoki:2008gv}
with new approaches being developed. These new methods involve the
calculation of the $CP$-odd $F_3(Q^2)$ form factor by treating the
$\theta$-parameter
perturbatively~\cite{Guadagnoli:2002nm,Facciolia:2004jz,Shintani:2005xg,Shintani:2014zra,Shindler:2015aqa}
or, simulating the theory with an imaginary
$\theta$~\cite{Aoki:2008gv,Guo:2015tla}. Alternative definitions of the
topological charge were also considered as, for example, replacing the
topological charge operator with the flavour-singlet pseudoscalar
density employing  the axial chiral Ward
identities~\cite{Guadagnoli:2002nm,Facciolia:2004jz}. 

Despite the recent progress, a lattice determination of the nEDM is
inherently difficult and still remains a challenging task. The
expectation value of an operator $\mc{O}$ in a theory 
with non-conserving  $CP$-symmetry can, in principle, be obtained by
using the path integral formulation, with the Lagrangian given in
\eq{eq:Lagrangian_Density}. Thus, the expectation value is given by
\bea
    \< \mc{O}(x_1,...,x_n) \>_{\theta} = \frac{1}{Z_\theta} \int d[U] d[\psi_f] 
    d[\bar{\psi}_f] ~ \mc{O}(x_1,...,x_n) ~ e^{-S_{\rm QCD}+i{\theta}
      \int  d^4x\, q(x)}\,,
    \label{eq:vev}
\eea
where $S_{\rm QCD} = \int d^4 x\, {\cal L}_{\rm QCD}$ and 
${\cal L}_{\rm QCD}$ is defined in \eq{eq:QCD_Lagrangian}. 
In what follows, we will use the notation $\< \dots \>_{\theta}$ to indicate
expectation values in the $CP$ non-conserving theory, where the
Chern-Simons term, ${\cal L}_{CS}$, is included. However, the
numerical determination of the expectation value given in
Eq.~(\ref{eq:vev}) suffers from the well-known sign problem due to the
imaginary character of ${\cal L}_{CS}$. Therefore, it is
not feasible to produce gauge configurations with the Lagrangian density of
\eq{eq:Lagrangian_Density} and carry out an adequate sampling of the
gauge field configuration space. We can overcome this
obstacle by treating the Chern-Simons contribution perturbatively
assuming that $\theta$ is a small parameter
\be
e^{i \theta \int d^4x \,q(x)} \equiv e^{i \theta {\cal Q}} = 1 + i \theta {\cal Q} + {\it O}(\theta^2)\,.
\ee
Thus, to leading order in $\theta$, we obtain 
\bea
  \< \mc{O}(x_1,...,x_n) \>_{\theta}= \left\< \mc{O}(x_1,...,x_n)
  \right\>_{\theta=0}+ i{\theta}\,\left\< \mc{O}(x_1,...,x_n) \left(
  \int  d^4x\, q(x) \right)\right\>_{\theta=0}+{O}(\theta^2)\,,
    \label{eq:vevtheta}
\eea
where
\be {\cal Q}=\int  d^4x \,q(x) \,,
\label{eq:topological_charge}
\ee
is the topological charge. This
expansion becomes our starting point for the calculation of the $CP$-odd
form factor $F_3(Q^2)$ and, consequently, of the nEDM.

In the remaining part of this section we present the methodology of
our work in order to extract $F_3$, based on the linear response of
the $CP$-violating strength parameter, $\theta$. We adopt the formulation
introduced in Ref.~\cite{Shintani:2005xg} as summarized below. 

In order to extract the $CP$-violating electric dipole form factor $F_3(Q^2)$
we consider the nucleon matrix element of
the electromagnetic current, given as
\bea
J_\mu^{\rm em} = \sum_f e_f\bar\psi_f  \gamma_\mu \psi_f\,,
\label{eq:Electromagnetic_Form_Factor}
\eea
where $e_f$ denotes the electric charge of the quark field
$\psi_f$. The nucleon matrix element of the electromagnetic operator
in the $\theta$-vacuum can be written as 
\be
{}_{\theta}\langle N ({\vec p}_f,s_f)\vert J_\mu^{\rm em}\vert N
({\vec p}_i,s_i)\rangle_{\theta}
= \bar u_N^\theta ({\vec p}_f,s_f)\, W_\mu^\theta (Q)\, u_N^\theta
({\vec p}_i,s_i)\,,
\label{eq:forms}
\ee
where $p_f$ ($p_i$) and $s_f$ ($s_i$) are the momentum and spin of the
final (initial) spin-$1/2$ nucleon state $N$. According to parity arguments, $W^\theta_\mu(Q)$
is decomposed into an even and an odd part. Up to order $\theta$, this has the form
\bea
W_\mu^\theta (Q)  &=& W_\mu^{\rm even}(Q)
+i\theta\ W_\mu^{\rm odd}(Q)\,.
\eea
The even part $W_\mu^{\rm even}(Q)$ can be written in terms of the $CP$-conserving Pauli and
Dirac form factors $F_1(Q^2)$ and $F_2(Q^2)$, respectively 
\be
W_\mu^{\rm even}(Q)= \gamma_\mu
F_1(Q^2) - i\,\frac{F_2(Q^2)}{2m_N}\,Q_\nu\,\sigma_{\nu\mu},
\label{eq:Weven}
\ee
while the odd part $W_\mu^{\rm odd}(Q)$ is written in terms of the electric dipole
$F_3(Q^2)$ and the anapole $F_A(Q^2)$ form factors
\be
W_\mu^{\rm odd}(Q)= -i\, \frac{F_3(Q^2)}{2m_N}\, Q_\nu\,\sigma_{\nu\mu}\gamma_5
+ F_A(Q^2)\left(Q_\mu \slashed{Q} - \gamma_\mu\,Q^2\right)\gamma_5\,.
\label{eq:Wodd}
\ee
In the absence of the $\theta$-term in the action, only the even part
remains. The additional form factors that 
arise in the odd part for a non-zero value of $\theta$ is
the $CP$-violating form factor, $F_3(Q^2)$, which gives the electric dipole
moment according to \eq{eq:dN}, and  the $P$-violating, but
$T$-preserving, form factor, $F_A(Q^2)$, which measures the anapole moment
of the nucleon. Hence, the anapole form factor is $C$-violating. Since the 
action is $C$-preserving such a form-factor is zero and, thus, will not be considered here.

One approach to study the electric dipole moment, is to generate gauge
configurations including the $\theta$-term in the action by considering 
an imaginary $\theta$~\cite{Guo:2015tla}. We instead use a real value of $\theta$ and
expand the matrix elements to leading order in $\theta$. This allows us to  make use of the 
gauge configurations generated without the $\theta$-term to evaluate
the appropriate expectation values, according to \eq{eq:vevtheta}. We
first examine the expressions with the $\theta$-term in the action before
we perform the expansion to leading order in $\theta$. We are interested in
the  three-point function   given by 
\be
G_{\rm 3pt}^{\mu, \left(\theta \right)} (\qvec,t_f,t,t_i) \equiv 
\langle J_N (\vec p_f,t_f) \, J_{\mu}^{\rm em}(\vec q, t)\, 
{\overline J}_N({\vec p_i},t_i) \rangle_{\theta}\,,
\label{eq:Definition_Three_Point_Function}
\ee
where ${\overline J}_N (\vec p_i,t_i)$ and $J_N (\vec p_f,t_f) $ are the
interpolating operators at the time-space of the source and the sink, respectively. Thus, according
to our notation the nucleon is created at time $t_i$ with momentum
$\vec{p_i}$, it couples with the electromagnetic current at some later time
$t$, and then is annihilated at time $t_f$ having momentum $\vec{p_f}$. 
The momentum transfer is thus $\vec{Q}=\vec{q}=\vec{p_f}-\vec{p_i}$.
The subscript $\theta$ in \eq{eq:Definition_Three_Point_Function} implies
that the expectation value is taken with respect to the action
including the $CP$-violating term (\eq{eq:vev}). \par

Inserting  complete set of states
in the three-point function of \eq{eq:Definition_Three_Point_Function}
one obtains 
\bea
G_{\rm 3pt}^{\mu, \left(\theta \right)} (\qvec, t_f, t_i, t ) & \simeq &
e^{-E^f_{N^\theta}(t_f-t)}e^{-E_{N^\theta}^i (t-t_i)}\nn
&\times&
\sum_{s_f,s_i}\langle J_N \vert N
({\vec p_f},s_f)\rangle_\theta
{}_\theta\langle N({\vec p_f},s_f)\vert J_\mu^{\rm em}\vert N
({\vec p_i},s_i)\rangle_\theta
{}_\theta\langle N({\vec p_i},s_i)\vert {\overline J_N} \rangle\,,
\label{eq:Three_Point_Function}
\eea
with $E_{N^\theta}^f {{\equiv E_{N^\theta}^f({\vec p}_f)}} =\sqrt{{\vec p}^{\ 2}_f+m_{N^\theta}^2}$ and
$E_{N^\theta}^i {{\equiv E_{N^\theta}^i({\vec p}_i)}} =\sqrt{{\vec p}_i^{\ 2}+m_{N^\theta}^2}$. The 
nucleon states for non-zero $\theta$-term are denoted as $|
N\rangle_\theta$, and are normalized as
\bea
\langle J_N \vert N ({\vec p},s)\rangle_\theta = Z_N^\theta
u_N^\theta (\vec p,s), \ \ \ {\rm and} \ \ \
_\theta\langle N ({\vec p},s)\vert  {\overline J_N} \rangle
=
(Z_N^\theta)^* \bar u_N^\theta ({\vec p},s)\,.
\eea
The spinors $u_N^\theta (\vec p,s)$ and $\bar u_N^\theta (\vec p,s)$ satisfy the Dirac equation
\bea
  \left(i \slashed{p} + m_{N^{\theta}}e^{-i 2 \alpha(\theta)\gamma_5} \right)
u_N^\theta({\vec p},s)&=&
\bar u_N^\theta({\vec p},s)\left(i\slashed{p} + m_{N^{\theta}}e^{-i
2 \alpha(\theta)\gamma_5} \right) =0\,,
\eea
with the phase $e^{-i 2 \alpha(\theta)\gamma_5}$ appearing in the mass
term due to the $CP$ breaking in the $\theta$-vacuum. This requirement
suggests that $\alpha(\theta)$ is odd in $\theta$, while
$m_{N^{\theta}}$ and $Z^{\theta}_N$ are even. In the presence of the
$\theta$-term, the spinor sum takes a phase in the mass term, becoming
\bea
\Lambda^{\left( \theta \right)}_{1/2}(\vec{p} \hspace{0.05cm} ) =  \sum_{s} u_N^\theta (\vec{p},s)\bar u_N^\theta (\vec{p},s)
= \frac{-i \slashed{p} + m_{N^\theta}e^{i
2 \alpha(\theta)\gamma_5}}{2E_{N^\theta}}\,.
\label{eq:2-pt_in_theta}
\eea

For small values of the $\theta$ parameter we can expand the
above expressions, and to leading order in $\theta$,
we obtain
\bea
\alpha(\theta) = \alpha^1 \theta + O(\theta^3), \quad 
m_{N^\theta} = m_N + O(\theta^2) \quad {\rm and} \quad 
Z^{\theta}_{N} = Z_N + O(\theta^2)\,.
\label{eq:alpha_and_mN}
\eea
Upon substituting the above expressions in \eq{eq:Three_Point_Function}
one obtains
\bea
G_{\rm 3pt}^{\mu, \left(\theta \right)} (\qvec, t_f, t_i, t ) &=& \vert Z_N\vert^2
e^{-E_{N}^f(t_f-t)}e^{-E_{N}^i (t-t_i)}
\frac{-i\slashed{p}_f + m_{N}(1+2 i \alpha^1\theta \gamma_5)}{2E^f_{N}} \nonumber \\
&\times& 
\left[W_\mu^{\rm even}(Q) +i\theta W_\mu^{\rm odd}(Q) \right]
\frac{-i\slashed{p}_i + m_{N}(1+2 i \alpha^1\theta
\gamma_5)}{2E_{N}^i}
+O(\theta^2)\,,
\label{eq:3pt_ff}
\eea
up to linear terms in $\theta$. 
When combined with the leading order of \eq{eq:vevtheta}, the above
equation contains all the information needed for the evaluation of $F_3(Q^2)$.
The advantage of this method is that the Green's functions of the
theory in the presence of the $CP$-violating term can be expressed in terms of
expectation values obtained using the gauge configurations generated
with the action with $\theta$ set to zero. For the three-point
function, this gives 
\bea
 G_{\rm 3pt}^{\mu, \left(\theta \right)} (\qvec, t_f, t_i, t ) & = &
 \langle J_N ({\vec p_f},t_f) J_\mu^{\rm em}(\qvec, t)
\overline J_N ({\vec p_i},t_i) \,\rangle_\theta \\
& = & G_{\rm 3pt}^{\mu,(0)} (\qvec, t_f, t_i, t ) +
i \,\theta \, G_{\rm 3pt, {\cal Q}}^{\mu,(0)} (\qvec, t_f, t_i, t ) +{\it O} \left( \theta^2 \right),
\label{eq:Three_point_perturbative}
\eea
where
\bea
G_{\rm 3pt}^{\mu,(0)} (\qvec, t_f, t_i, t )&=&\langle J_N({\vec p}_f,t_f) J^{\rm em}_\mu(\vec q, t)
 {\overline J}_N({\vec p}_i,t_i)\rangle, \\
G_{\rm 3pt, {\cal Q}}^{\mu, \left(0\right)} (\qvec, t_f, t_i, t )&=&\langle J_N ({\vec p}_f,t_f) J^{\rm em}_\mu(\vec q, t) 
{\overline J}_N({\vec p}_i,t_i) {\cal Q} \rangle.
\label{eq:Definition_of_three_point_functions}
\eea
We now equate \eq{eq:Three_point_perturbative} to \eq{eq:3pt_ff} and
express $G_{\rm 3pt}^{\mu,(0)}$ and 
$G_{\rm 3pt, {\cal Q}}^{\mu, \left(0\right)}$ through the $W_\mu^{\rm
  even}(Q)$ and $W_\mu^{\rm odd}(Q)$ parts, respectively 
\bea
G^{(0)}_{\rm 3pt} (\qvec, t_f, t_i, t ) &=& \vert Z_N\vert^2
e^{-E_{N}^f (t_f-t)}e^{-E_{N}^i (t-t_i)}
\frac{-i\slashed{p}_f + m_{N}}{2E_{N}^f} W_\mu^{\rm even}(Q)
\frac{-i\slashed{p}_i + m_{N}}{2E_{N}^i},
\label{eq:3pt_even_1}\\
G^{(0)}_{{\rm 3pt}, {\cal Q}} (\qvec, t_f, t_i, t ) &=&  \vert Z_N\vert^2
e^{-E_{N}^f(t_f - t)}e^{-E_{N}^i (t-t_i)}
\Bigl[
\frac{-i\slashed{p}_f + m_{N}}{2E_{N}^f} W_\mu^{\rm odd}(Q)
\frac{-i\slashed{p}_i + m_{N}}{2E_{N}^i} \nonumber \\ 
&+&
\frac{2 \alpha^1m_N}{2E_N^f}\gamma_5 W_\mu^{\rm even}(Q)
\frac{-i\slashed{p}_i + m_{N}}{2E_{N}^i} + \frac{-i\slashed{p}_f + m_{N}}{2E_{N}^f} W_\mu^{\rm even}(Q)
\frac{2 \alpha^1m_N}{2E_N^i}\gamma_5
\Bigr].\label{eq:3pt_odd_1}
\eea
A similar analysis can be carried out for the case of the two-point
functions. 
In the presence of the $\theta$-term of \eq{eq:Lagrangian_Density} the
relevant two-point function is given by 
\bea
G_{\rm 2pt}^{\left(\theta \right)} (\qvec,t_f, t_i) &\equiv & \langle  J_N({\vec q},t_f) {\overline J}_N({\vec q},t_i) \rangle_{\theta} = \vert Z_N^\theta\vert^2 e^{-E_{N^\theta} (t_f-t_i)}\frac{-i\slashed{Q} +
m_{N^\theta}
e^{i 2 \alpha(\theta)\gamma_5}}{2E_{N^\theta}}\,. 
\label{eq:2ptOperator}
\eea
By treating the Chern-Simons term perturbatively one obtains
\be
G_{\rm 2pt}^{\left(\theta \right)}(\qvec,t_f, t_i)  = 
G_{\rm 2pt}^{\left(0 \right)}(\qvec,t_f, t_i) + i\, \theta\,
G_{\rm 2pt,{\cal Q}}^{\left(0 \right)}(\qvec,t_f, t_i) + O \left( \theta^2 \right),
\label{eq:2ptPathIntegral}
\ee
where
\bea
G_{\rm 2pt}^{\left(0 \right)}(\qvec,t_f, t_i) \quad = &\langle  J_N(\vec{q},t_f) {\overline J}_N(\vec{q},t_i) \rangle & = \quad
\vert Z_N\vert^2 e^{-E_{N} t}\frac{-i\slashed{Q} + m_{N}}{2E_{N}} \,,\\
 G^{\left(0\right)}_{{\rm 2pt}, {\cal Q}}(\vec{q},t_f, t_i) \quad = & \langle  J_N(\vec{q},t_f) {\overline J}_N(\vec{q},t_i) {\cal Q} \rangle & = \quad
\vert Z_N\vert^2 e^{-E_{N} t}\frac{ 2 \alpha^1 m_{N}}{2E_{N}} \gamma_5\,.
\label{eq:2pt_Q}
\eea

From the two-point function $ G^{\left(0\right)}_{{\rm 2pt}, {\cal Q}}$
one can extract the parameter $\alpha^1$, which enters in the
decomposition  of the nucleon matrix element to leading order in
$\theta$. This will be further discussed in Section~\ref{sec:Plateau_method_to_extract_alpha}. \par

Using \eq{eq:2pt_Q} in conjunction with  Eqs.~(\ref{eq:3pt_even_1})
and (\ref{eq:3pt_odd_1})  one can obtain $F_3(Q^2)$ from the leading
order in $\theta$. To summarize, this approach requires: i) the evaluation of
two- and three-point functions  using gauge configurations simulated by
setting $\theta=0$ , ii) the computation of the topological charge ${\cal Q}$
that will be explained in \sec{sec:The_Topological_Charge}, and iii)
choosing appropriate projectors in order to extract $F_3(Q^2)$ using
suitable ratios of correlation functions, as will be explained in
Section~\ref{sec:Plateau_method_to_extract_F_3}. \par

\section{Correlation functions}
\label{sec:correlators}
Taking into account the Dirac structure of the matrix
elements the two- and three-point functions are expressed as
\bea
G_{\rm 2pt}(\vec{q},t_f,t_i,{{\Gamma_0}})  &\equiv & 
\vert Z_N\vert^2 e^{-E_{N} (t_f-t_i)} {{\Gamma_0}}^{\alpha \beta} \Bigl[{\Lambda_{1/2}}(\vec{q})\Bigr]_{\alpha \beta}\, ,  \label{eq:2pt_even} \\
G_{\rm 2pt,{\cal Q}} (\vec{q},t_f,t_i,{\Gamma_5}) &\equiv & 
\vert Z_N\vert^2 e^{-E_{N} (t_f-t_i)} {\Gamma_5}^{\alpha \beta}
\Bigl[\frac{\alpha^1 m_{N}}{E_{N}} \gamma_5 \Bigr]_{\alpha \beta}\,, \label{eq:2pt_odd} \\
G^\mu_{\rm 3pt} (\qvec, t_f, t_i, t ,\Gamma_k) &=& \vert Z_N\vert^2
e^{-E_{N}^f(t_f-t)}e^{-E_{N}^i (t-t_i)}
\Gamma_k^{\alpha \beta} \Bigl[ {\Lambda_{1/2}}(\vec{p}_f) W_\mu^{\rm even}(Q)
{\Lambda_{1/2}}(\vec{p}_i) \Bigr]_{\alpha \beta}\,,
\label{eq:3pt_even_2}\\
G^\mu_{\rm 3pt, {\cal Q}} (\qvec, t_f, t_i, t ,\Gamma_k) &=&  \vert Z_N\vert^2
e^{-E_{N}^f(t_f-t)}e^{-E_{N}^i (t-t_i)}
\Gamma_k^{\alpha \beta} \Biggl[
 {\Lambda_{1/2}}(\vec{p}_f) W_\mu^{\rm odd}(Q)
{\Lambda_{1/2}}(\vec{p}_i) \nonumber \\ 
&+&
\frac{\alpha^1m_N}{E_N^f}\gamma_5 W_\mu^{\rm even}(Q)
{\Lambda_{1/2}}(\vec{p}_i)+ {\Lambda_{1/2}}(\vec{p}_f) W_\mu^{\rm even}(Q)
\frac{\alpha^1m_N}{E_N^i}\gamma_5
\Biggr]_{\alpha \beta}\,.
\label{eq:3pt_odd}
\eea
Note that we dropped the $\theta$-superscript on the two- and
three-point functions, since from now on we consider expectation
values with $\theta=0$. Also we define $\Lambda_{1/2}$
that is given in \eq{eq:2-pt_in_theta} by setting $\theta=0$ and
$\Gamma_j$ is an appropriate  projector acting on the Dirac structure
of the two- and three-point functions. For the three-point function we
choose the three projectors given by
\bea
\Gamma_k=\frac{i}{4}(\idnty+\gamma_0)\,\gamma_5\,\gamma_k  \quad (k=1,2,3)\,,
\label{eq:three_point_function}
\eea
which can disentangle the four form factors.
Using three projectors (one for each spatial direction) instead of
summing over $k$ increases  the computational cost, since separate
sequential inversions are required for each projector. We use inexact
deflation~\cite{Stathopoulos:2007zi,Stathopoulos:2009zz} to speedup
the inversions. In total we do three inversions for the projectors and
four for each sink-source time separations. The use of inexact deflation 
brings a factor of three speedup enabling us to do twelve inversions
with the cost of four~\cite{Abdel-Rehim:2013wlz}. The use of the three
projectors is needed for the application of the position space methods
for the extraction of $F_3(0)$ as discussed in Subsection~\ref{sec:y_summation}.

Two projectors are employed in the evaluation of the two-point functions, namely
\bea
\Gamma_0 = \frac{1}{4} \left( \idnty+\gamma_0 \right)\,, \quad\Gamma_5=\frac{\gamma_5}{4}\,,
\label{eq:two_point_function}
\eea
with the projector $\Gamma_5$ needed in order to  access  the parameter $\alpha^1$.

\begin{figure}[htb]
\vspace{-3cm}
\centerline{\hspace{0.0cm}\includegraphics[scale=0.6,angle=270]{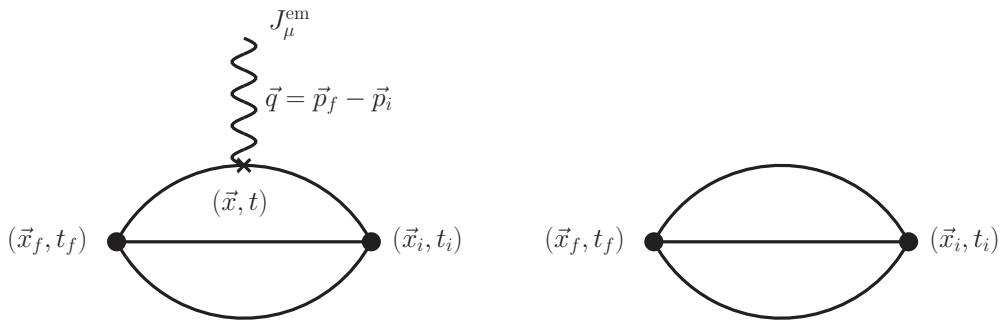}}
\vspace{-5cm}
\caption{\label{fig:3pt_2pt_diagrams} Diagrammatic representation of
the connected three-point function (left) and two-point function (right).}
\end{figure}

In the actual computation, we employ the proton interpolating operators, which in the physical basis reads
\be
{J}_N(x) {=} \epsilon^{abc} u^a_\alpha \left[ u(x)^{\top b} \C\gamma_5 d(x)^c\right]\,,
\ee
where $\C$ is the charge conjugation matrix. Since we employ
degenerate up and down quarks, the proton Green's functions are
equivalent to those of the neutron. Our framework preserves isospin so the
proton and neutron electric dipole moments are the same up to a sign. We use Gaussian smeared
quark fields~\cite{Alexandrou:1992ti,Gusken:1989qx} to increase the
overlap with the neutron state and decrease the overlap with excited
states. The smeared interpolating fields are given by
\bea
q_{\rm smear}^a(t,\vec x) &=& \sum_{\vec y} F^{ab}(\vec x,\vec y;U(t))\ q^b(t,\vec y)\,,\\
F &=& (\idnty + {a_G} H)^{N_G} \,, \nonumber\\
H(\vec x,\vec y; U(t)) &=& \sum_{i=1}^3[U_i(x) \delta_{x,y-\hat\imath} + U_i^\dagger(x-\hat\imath) \delta_{x,y+\hat\imath}]\,. \nonumber
\eea
In addition, we apply APE-smearing to the gauge fields $U_\mu$
entering the hopping matrix $H$. The parameters for the Gaussian
smearing $a_G$ and $N_G$ are optimized using the nucleon ground
state~\cite{Alexandrou:2008tn}. Different combinations of $N_{\rm G}$
and $a_{\rm G}$, have been tested in previous work and it was found
that combinations of $N_G$ and $a_G$ that give a root mean square
radius of about $0.5$~fm are optimal for suppressing excited states in
the nucleon case. The results of this work have been produced with
$N_G=50,\, a_G=4,\,N_{\rm APE}=20,\,a_{\rm APE}=0.5$.

Besides the connected three-point function depicted in
Fig.~\ref{fig:3pt_2pt_diagrams} there is a  disconnected diagram that
can contribute to the nucleon matrix element of the electromagnetic
current. However, from a previous study~\cite{Abdel-Rehim:2013wlz}, we found for $\sigma_{\pi N}$ a maximum disconnected contribution of about 10\% of the connected while for the electromagnetic form factors at the lowest available  $Q^2$ contributions are less than 1\%. Hence, although disconnected contributions to $F_3$ have not yet been studied, we expect them  to be of similar magnitude as for the other nucleon form factors. Since the nEDM is a rather noisy observable we  neglect the disconnected contribution 
in the present computation. Hence, the correlators of Eqs.~(\ref{eq:3pt_even_2}) - (\ref{eq:3pt_odd}) are  calculated using the
connected diagram only.  For the computation of the connected three-point function  we employ sequential inversions through the sink. 

In our computation we take the nucleon creation operator at a fixed position
$\vec{x}_i=\vec{0}$ (source) with momentum ${\vec p_i}$. The
nucleon annihilation operator at a later time $t_f$ (sink) carries
zero momentum, i.e. we set $\vec{p}_f=0$. The electromagnetic current
$J^{\rm em}_{\mu}$ couples to a quark at an intermediate time $t$
(insertion) and carries momentum $\vec{q}$ while translation
invariance enforces $\vec{q}=-\vec{p}_i$.  At a fixed sink-source time
separation, $t_{\rm sep}=t_f-t_i$, we obtain results for all possible momentum
transfers and insertion times $t$ with one set of sequential inversions
per choice of $t_f$. We consider three values of $t_{\rm sep}$ in order to check for
ground state dominance.

\section{Extraction of $\alpha^1$}
\label{sec:Plateau_method_to_extract_alpha}
The $\alpha^1$ parameter can be determined from a ratio of  two-point
functions with the appropriate projectors. Although, there is more than one
choice for extracting $\alpha^1$, we find that the optimal choice with respect to the resulting signal-to-noise ratio is given by
\bea
{\rm R}_{\rm 2pt}(\alpha^1,t_f,t_i) = \frac{G_{\rm 2pt, {\cal Q}}
  \left(0, t_f, t_i,\Gamma_5 \right) }{G_{\rm 2pt} \left(0, t_f, t_i, \Gamma_0 \right)}\,,
\label{eq:2_pt_function}
\eea
with the two-point functions defined in Eqs.~(\ref{eq:2pt_even}) -
(\ref{eq:2pt_odd}). For the extraction of $\alpha^1$ no momentum is
required and we thus set $\vec{q}=0$. By taking the large $t_{\rm sep}$ limit 
\eq{eq:2_pt_function} results in a time-independent quantity
(plateau) 
\bea
\Pi_{\rm 2pt} (\alpha^1)= \lim_{t_{\rm sep} \to \infty} {\rm R}_{\rm 2pt}(\alpha^1,t_f,t_i)= \alpha^1\,,
\eea
which  can be fitted to a constant yielding  $\alpha^1$. The determination of
$\alpha^1$ requires the evaluation of the topological charge, as indicated by the subscript ${\cal Q}$ on the two-point function of
\eq{eq:2_pt_function}. The dependence of $\alpha^1$ on the definition of ${\cal Q}$ is investigated  in Section~\ref{sec:calculation_of_alpha}. 

\section{Extraction of $F_3(Q^2)$}
\label{sec:Plateau_method_to_extract_F_3}
Before we discuss the calculation of the topological charge ${\cal Q}$
needed for both the evaluation of $\alpha^1$ and $F_3(Q^2)$ we first explain our methods to
extract $F_3(Q^2)$ at $Q^2=0$ using the three- and two-point
functions mentioned in the previous section.

Our calculation proceeds by evaluating the diagrams of
\fig{fig:3pt_2pt_diagrams}. In order to cancel all unknown overlaps
and the exponential time dependence we take  appropriate ratios
using the two- and three-point functions of \eq{eq:2pt_even} and
\eq{eq:3pt_odd}. Specific linear combinations for three-point functions 
given in \eq{eq:3pt_even_2}
 are also constructed in order to eliminate the dominant form factors 
$F_1$ and $F_2$, as explained below. When the insertion-source and
sink-insertion time separations are large enough so that contamination
from higher excitations is small we obtain a time-independent quantity
(plateau) to which we fit to extract $F_3(Q^2)$. The traces of
Eqs.~(\ref{eq:2pt_even}) - (\ref{eq:3pt_odd}) are calculated using Dirac
trace algebra, which is implemented with a symbolic analysis package
in Mathematica (see, e.g., Ref.~\cite{Constantinou:2009tr}). 
 For
the evaluation of $F_3$ we consider the following ratio:
\bea
{\rm R}^{\mu}_{\rm 3pt,{\cal Q}} \left(\vec{q}, t_f, t_i, t,\Gamma_k \right) &
= & \frac{G^{\mu}_{\rm 3pt, \cal Q} (\vec{q}, t_f, t_i, t,
  \Gamma_k)}{G_{\rm 2pt}(\vec{q},t_f-t_i,\Gamma_0)} \nonumber \\
& \times & \sqrt{\frac{G_{\rm 2pt}(\vec{q},t_f-t,\Gamma_0)G_{\rm
      2pt}(\vec{0},t-t_i,\Gamma_0)G_{\rm
      2pt}(\vec{0},t_f-t_i,\Gamma_0)}{G_{\rm 2pt}(\vec{0},t_f-t,\Gamma_0)G_{\rm 2pt}(\vec{q},t-t_i,\Gamma_0)G_{\rm 2pt}(\vec{q},t_f-t_i,\Gamma_0)}}\,,
\label{eq:Ratio_3pt_2pt}
\eea
for each one of the three projectors given in
\eq{eq:two_point_function}. For sufficiently large separations $t_f-t$
and $t-t_i$ these ratios become time-independent (plateau region)
\bea
\Pi^{\mu}_{\rm 3pt,{\cal Q}} \left( \Gamma_k \right)= \lim_{t_f-t \to
  \infty} \lim_{t-t_i \to \infty} {\rm R}^{\mu}_{\rm 3pt,{\cal Q}} \left(\vec{q}, t_f, t_i, t,\Gamma_k \right)\,,
\eea
where $\mu$ is the current index.
Using \eq{eq:2pt_even} and \eq{eq:3pt_odd},  carrying out
the Dirac algebra  and simplifying using our kinematics, we
obtain the following expressions: 
\bea
\Pi^{0}_{\rm 3pt,{\cal Q}} \left( \Gamma_k \right) & = & i\,C Q_k \left[ \frac{ \alpha^1 F_1(Q^2)}{2 m_N} + \frac{(E_N+3 m_N ) \alpha^1 F_2(Q^2)}{4 m^2_N} + \frac{ (E_N+m_N) F_3(Q^2)}{ 4m^2_N} \right]\,, \label{eq:plateau_P0}\\
\Pi^{j}_{\rm 3pt,{\cal Q}} \left( \Gamma_k \right) & = & C  \Biggl[
  \frac{ (E_N-m_N)\alpha^1 \delta_{k,j} F_1(Q^2)}{2 m_N} + \frac{Q_k
    Q_j F_3(Q^2)}{4 m^2_N} \nonumber \\ 
&+&\frac{ \alpha^1 F_2(Q^2) \left( \left( 2 E_N m_N - 2 m_N^2 \right)
    \delta_{k,j}  +  Q_k Q_j \right)}{4 m^2_N}
  \Biggl]\quad {{j=1,2,3}}\,, \label{eq:plateau_Pj} 
\eea
where $C=\sqrt{\frac{2m_N^2}{E_N\left(E_N+m_N\right)}}$. 

The presence of the $CP$-even form factors in
Eqs.~(\ref{eq:plateau_P0}) - (\ref{eq:plateau_Pj}) requires the values of  $F_1(Q^2)$ and $F_2(Q^2)$. 
We can eliminate
$F_1$ and $F_2$ by substituting them with the appropriate $CP$-even ratios 
in large time limit~\cite{Alexandrou:2013joa}. These ratios are given by
\bea
\label{eq:plateau_F1}
Q_i\,F_1(Q^2) & = & \frac{1}{C (E_N+m_N)} \Biggl[4\,i\,m_N^2\, \Pi^{i}_{\rm 3pt}(\Gamma_0) + 
m_N\,(E_N-m_N)\,\epsilon^{ijk}\,\Pi^{j}_{\rm 3pt}(\Gamma_k) \Biggr]\,,\\
\label{eq:plateau_F2}
Q_i\,F_2(Q^2) & = & \frac{1}{C (E_N+m_N)} \Biggl[-4\,i\,m^2_N\,\Pi^{i}_{\rm 3pt}(\Gamma_0) +
2\,m_N^2\,\epsilon^{ijk}\,\Pi^{j}_{\rm 3pt}(\Gamma_k) \Biggr]\,,
\eea
where  a summation over the spatial indices $j,k$ is implied. Thus, we can
extract $F_3$ from the following combination
\bea
\label{eq:plateau_F3}
Q_i\,F_3(Q^2) &=& \frac{1}{C(E_N+m_N)} \Biggl[  4\,m_N^2\,\Pi^{0}_{\rm 3pt, \cal{Q}}(\Gamma_i) \nonumber \\
& &-\alpha^1\,2m_N\left(4\,i\,m_N^3\, \Pi^{i}_{\rm 3pt}(\Gamma_0) + 
m_N\,(E_N-m_N)\,\epsilon^{ijk}\,\Pi^{j}_{\rm 3pt}(\Gamma_k) \right)\nonumber\\
& &  -\alpha^1 (E_N+3\,m_N)\left(-4\,i\,m^2_N\,\Pi^{i}_{\rm 3pt}(\Gamma_0) +
2\,m_N^2\,\epsilon^{ijk}\,\Pi^{j}_{\rm 3pt}(\Gamma_k)\right)\Biggr]\,,
\eea
where, instead of $F_1$ and $F_2$, we substitute
Eqs.~(\ref{eq:plateau_F1}) - (\ref{eq:plateau_F2}) in \eq{eq:plateau_P0} allowing us to extract $F_3(Q^2)$. We first calculate $\alpha^1$ and  then take the linear combination of $\Pi^{0}_{\rm 3pt, \cal{Q}}(\Gamma_k)$, $\Pi^{i}_{\rm 3pt}(\Gamma_0)$ and $\Pi^{j}_{\rm 3pt}(\Gamma_k)$ from which the plateau value directly yields $Q_i\,F_3(Q^2)$.  An analysis using separately the ratios involving $F_1$, $F_2$ and
$F_3$ (Eqs.~(\ref{eq:plateau_P0}), (\ref{eq:plateau_F1}),
(\ref{eq:plateau_F2})) has also been carried out using singular value
decomposition, which gave consistent numerical results for $F_3(Q^2)$. Although both
Eqs.~(\ref{eq:plateau_P0}) - (\ref{eq:plateau_Pj}) involve
$F_3$, we use Eq.~(\ref{eq:plateau_P0}), which results in a  better signal-to-noise ratio. Moreover, the absence of the momentum factor $Q_i$ in front of $F_1$ and $F_2$ in \eq{eq:plateau_Pj} does not allow the usage of \eq{eq:plateau_F1} and \eq{eq:plateau_F2} to eliminate them in favor of $F_3$. The case $j\ne k$ does not allow access to the lowest momentum transfer making the extrapolation to $Q^2=0$ less reliable.

A more convenient, but equivalent, procedure is to introduce the
electric and magnetic Sachs form factors, $G_E$ and $G_M$, instead of
the Dirac and Pauli. These are related via
\begin{align}
 G_E\l(Q^2\r) &= F_1\l(Q^2\r) - \frac{Q^2}{4m_N^2} F_2\l(Q^2\r) \,, \\
 G_M\l(Q^2\r) &= F_1\l(Q^2\r) + F_2\l(Q^2\r) \,.
\end{align}
Employing an appropriate choice of projectors and insertion indices we find
\begin{align}
 \Pi^{j}_\mathrm{3pt}\l(\Gamma_0\r) &= -C\frac{i}{2\,m_N} Q_j G_E\l(Q^2\r) \label{eq:q_G_E} \,, \\
 \Pi^{j}_\mathrm{3pt}\l(\Gamma_k\r) &= -C\frac{1}{2\,m_N} \epsilon_{jik} Q_i G_M\l(Q^2\r) \label{eq:q_G_M} \,,
\end{align}
with the indices $i,j,k$ being spatial.
By combining Eqs.~(\ref{eq:q_G_E}) - (\ref{eq:q_G_M}) with
\eq{eq:plateau_P0} we extract the following linear combination of
ratios 
\begin{equation}
 \Pi^k_{F_3} = -i \Pi^{0}_{\rm 3pt,\cal Q}(\Gamma_k) + 
i\,\alpha^1\,\Pi^{k}_\mathrm{3pt}\l(\Gamma_0\r) + 
\alpha^1\,\frac{1}{2}\sum_{i,j=1}^3\epsilon_{jki}\Pi^{j}_\mathrm{3pt}\l(\Gamma_i\r) =
\frac{C (E_N + m_N)}{4\,m_N^2}\,Q_k\, F_3(Q^2)\,,
\label{eq:plateau_F3_comb}
\end{equation}
for which the decomposition only depends on the desired form factor
$F_3(Q^2)$.

As can be seen from~\eq{eq:plateau_P0}, the appearance of the momentum
transfer as a multiplicative factor in front of $F_3(Q^2)$ does not
allow the calculation of $F_3(Q^2=0)$ directly in momentum
space. We explain below three different techniques that allow
us to extract $F_3(0)$.\par 

\smallskip
\subsection{Extraction of $F_3(0)$ through a dipole fit}

The first approach is the commonly used parameterization of the $Q^2$-dependence
of  $F_3(Q^2)$  extracted  from \eq{eq:plateau_P0}
followed by  a fit to extract $F_3(0)$ treating it as  a fitting parameter. We use a dipole Ansatz~\cite{Shindler:2015aqa} for $Q^2$-dependence of $F_3(Q^2)$ of the form 
\bea
F_3 \left( Q^2 \right) = \frac{F_3(0)}{\left( 1 + \displaystyle\frac{Q^2}{m_{F_3}^2} \right)^2}\,,
\label{eq:dipole_fit}
\eea
where $F_3 \left(0\right)$ and $m_{F_3}$ are fit parameters.

\subsection{Position space methods}
\label{sec:position_space_methods}
The other approach that was recently developed and applied in the
study of the Pauli form factor $F_2(Q^2)$~\cite{Alexandrou:2014exa} is
based on removing the momentum factor in front of $F_3(Q^2)$ by employing the
so called position space methods. There are two ways to accomplish this: the first is to 
apply a continuum-like derivative to the ratio
and the second is  to first  determine the
plateau values in momentum space, then take the Fourier transform to
coordinate space and finally transform back to momentum space using
a continuous Fourier transform in such a way that the hindering
momentum factor is avoided in the final result. In what follows we will 
refer to  first position
space method  as ``application of the derivative to the ratio'' 
whereas to the
second  as  ``elimination of the momentum in the plateau region''. We briefly
explain these techniques in the next two subsections. \par

\subsubsection{Application of the derivative to the ratio technique}
\label{sec:continuum_derivative_technique}

Assuming continuous momenta one can formally remove the 
$Q_k$ dependence in front of $F_3(Q^2)$ in {{\eq{eq:plateau_F3_comb}}} by
applying a derivative with respect to $Q_j$ 

\bea
\lim_{Q^2 \to 0} \frac{\partial}{\partial Q_j} \Pi^k_{F_3} ( \vec{Q} ) = \frac{C \left( E_N +m_N \right)}{4 m_N^2} \delta_{kj}  F_3(0)\,.
\label{eq:Continuum_derivative_plateau}
\eea
For simplicity we explicitly show the application of the derivative to the ratio in \eq{eq:Ratio_3pt_2pt}, which leads to the first term in~\eq{eq:plateau_F3_comb}; the generalization on the other two ratios is straightforward. This gives
\bea
\lim_{Q^2 \to 0} \frac{\partial}{\partial Q_j}  {\rm R}^{\mu}_{\rm
  3pt,{\cal Q}} \left(\vec{q}, t_f, t_i, t,\Gamma_k \right) & = & \lim_{Q^2 \to
  0}   \frac{\frac{\partial}{\partial Q_j}G^\mu_{{\rm 3pt}, {\cal Q}}
    (\vec{q}, t_f, t_i, t, \Gamma_k)}{G_{\rm 2pt}(\vec{q},t_f,t_i,\Gamma_0)}\,, \label{eq:ratio_fourier} \\ 
& = & \frac{1}{G_{\rm 2pt}(\vec{0},t_f,t_i,\Gamma_0)} \cdot  \hspace{-0.2cm} \sum_{x_j = -L/2 + a}^{L/2 -
    a} \left( \sum_{\substack{x_i=0 \\ i \ne j}}^{L -a} \hspace{-0.05cm}i x_j G^\mu_{\rm 3pt, {\cal Q}} (\vec{x}, t_f, t_i, t, \Gamma_k) \right) \label{eq:fourier},
\eea
where in the second line the three-point function $ G^\mu_{{\rm 3pt},
  {\cal Q}} (\vec{x}, t_f, t_i, t, \Gamma_k)$ is expressed in position
space. The derivative only acts on the three-point function since  any
derivatives acting on the two-point functions in the above expression
vanish exactly when setting $Q^2=0$. In finite volume this expression
approximates the derivative of a $\delta$-distribution in momentum
space, 
\bea
a^3 \hspace{-0.4cm}\sum_{x_j = -L/2 + a}^{L/2 - a} \hspace{-0.1cm} \left( \sum_{\substack{x_i=0 \\ i \ne j}}^{L -a} \hspace{-0.1cm} i x_j G^\mu_{\rm 3pt, {\cal Q}} (\vec{x}, t_f, t_i,
t, \Gamma_k) \right) \hspace{-0.2cm} &=&  \hspace{-0.2cm} {\frac{1}{V} \sum_{\vec{k}}  \hspace{-0.1cm} \left( a^3 \hspace{-0.4cm}\sum_{x_j = -L/2 + a}^{L/2 - a}  \hspace{-0.1cm} \left( \sum_{\substack{x_i=0 \\ i \ne j}}^{L -a} i
x_j {\rm exp} \left( i \vec{k} \vec{x}  \right)\right)   \right)G^\mu_{{\rm 3pt}, {\cal Q}} (\vec{k}, t_f, t_i, t, \Gamma_k)\,,} \nonumber \\
& & \stackrel{L\rightarrow \infty}{\longrightarrow}\frac{1}{\left( 2 \pi
  \right)^3} \int d^3 \vec{k} \frac{\partial}{\partial k_j}
\delta^{(3)}(\vec{k})\, G^\mu_{{\rm 3pt}, {\cal Q}} (\vec{k}, t_f, t_i, t, \Gamma_k). \label{eq:summation}   
\eea
For finite $L$ this implies a residual $t$-dependence $G^\mu_{{\rm 3pt}, {\cal Q}} (\vec{q}, t_f, t_i, t, \Gamma_k)
\sim {\rm exp} \left( -\Delta E_N t  \right)$ with $\Delta E_N = E_N \left( \vec{q} \right) - m_N$. Only  for $L \to \infty$ we have
$\Delta E_N \to 0$. \par 

According to the above formulation, the basic building blocks for this
technique are the standard two-point functions and the derivative-like three point functions ${\partial}G^\mu_{{\rm 3pt},
  {\cal Q}} (\vec{q}, t_f, t_i, t, \Gamma_k)/{\partial Q_j}$. In the
actual lattice computation this involves the calculation of the full
three-point function in position space before multiplying by $x_j$,
taking the Fourier transformation and forming the ratio of
\eq{eq:fourier}. Additionally, this technique requires a large enough
lattice extent $L$ for the summation in \eq{eq:summation} to approximate
the $\delta$-function.

 In order to extract $F_3(0)$ from the decomposition in
Eq.~(\ref{eq:plateau_F3_comb}) one performs the derivative on the other two three-point functions following the same procedure as outlined above.
Namely, one needs the derivatives  
${\partial}G^\mu_{{\rm 3pt}, {\cal Q}} (\vec{q}, t_f, t_i, t, \Gamma_k)/{\partial Q_j}$, 
${\partial}G^k_{{\rm 3pt}} (\vec{q}, t_f, t_i, t, \Gamma_0)/{\partial Q_j}$ as well as
${\partial}G^k_{{\rm 3pt}} (\vec{q}, t_f, t_i, t, \Gamma_i)/{\partial Q_j}$. 
The residual $t$-dependence that may remain in
\eq{eq:Continuum_derivative_plateau} of the form
 ${\rm exp} \left( -\Delta E_N t  \right)$ is expected to vanish only as $L \to \infty$.

\subsubsection{The elimination of the momentum in the plateau region technique}
\label{sec:y_summation}

This method was originally developed for the nucleon isovector
magnetic moment or equivalently $F_2(0)$~\cite{Alexandrou:2014exa} and
allows to extract $F_3(0)$ in a model-independent way without the
residual time dependence of the previous method. In principle, it is
not restricted to the case of a simple momentum prefactor, but can be
used to remove any kinematic structure that would otherwise prevent
the extraction of a form factor at zero momentum without making a fit
Ansatz. In the following we discuss the application of this method for
$F_3(0)$.
We want to stress that a key element in the extraction of $F_3(Q^2)$ using
this method is the use of the projectors of
\eq{eq:three_point_function} without summing over the index $k$. 

Our starting point is \eq{eq:plateau_F3_comb}
which as explained is obtained by combining the corresponding expression
for the $CP$-odd ratio in Eq.~(\ref{eq:plateau_P0}) with  Eqs.~(\ref{eq:q_G_E}) - (\ref{eq:q_G_M}).
As far as the $Q_k$-dependence is concerned \eq{eq:plateau_F3_comb} is now
similar to the one for the magnetic form factor in
Eq.~(\ref{eq:q_G_M}). Therefore, we adopt the elimination of the momentum in the plateau region
technique as discussed in Ref.~\cite{Alexandrou:2014exa} for this
particular linear combination of ratios to obtain a continuous curve
for $F_3(Q^2)$ from which the value of $F_3(0)$, and consequently the
nEDM, can be extracted. In the following we briefly outline the basic
idea behind the elimination of the momentum in the plateau region technique referring the reader to
Ref.~\cite{Alexandrou:2015dc} for more details. \par 

While for the application of the derivative to the ratio approach the time-dependence is
only fitted after applying the derivative in position space, the
elimination of the momentum in the plateau region technique aims at removing any time-dependence
before applying the derivative. For now we restrict ourselves to
on-axis momenta, e.g. ${\vec{q}}=(\pm Q_1, 0,0)^T$ (and all permutations
thereof). After forming the combination of ratios in
Eq.~(\ref{eq:plateau_F3_comb}) we average over all momentum directions
taking only index combinations into account that give a non-zero
contribution for a given $Q_1$-value. The resulting, averaged ratios are
denoted by $\Pi(Q_1)$. \par 

Applying a Fourier transform on $\Pi(Q_1)$ gives the corresponding ratio
$\Pi(y)$ in position space, which satisfies $\Pi(y) \approx -\Pi(-y)$
up to statistical fluctuations. 
Note that in an actual lattice simulation the Fourier transform requires an additional cutoff ${Q_1}_\mathrm{max}$, because the calculation of $\Pi({Q_1})$ will be restricted to a limited number of lattice momenta due to noise. Typically this cutoff is much smaller than the maximally allowed lattice momentum. Since we have 
\begin{equation}
 \Pi(y)=\l\{\begin{array}{ll}
  +\Pi(n), & n=0,...,N/2\,, \\
  -\Pi(N-n), & n=N/2+1,...,N-1, \ N=L/a\,,
 \end{array}\r. \,
\end{equation}
where $n=y/a$, we can average over positive and negative values of
$y$, yielding an exactly antisymmetric expression
$\overline{\Pi}(n)$. The most crucial part of this method is to
transform $\overline{\Pi}(n)$ back in a way that allows to introduce
continuous momenta. This can be achieved by rewriting the
corresponding Fourier transform in the following way 
\begin{align} 
 \Pi(k) &= \l[\exp(ikn)\overline{\Pi}(n)\r]_{n=0,\,N/2} + \sum\limits_{n=1}^{N/2-1}\exp(ikn)\overline{\Pi}(n) + \sum\limits_{n=N-1}^{N/2+1}\exp(ik(N-n))\overline{\Pi}(n)\,, \notag \\
        &=\l[\exp(ikn)\overline{\Pi}(n)\r]_{n=0,\,N/2} + 2i \sum\limits_{n=1}^{N/2-1} \overline{\Pi}(n) \sin\l(\frac{k}{2}\cdot (2n) \r) \,,
\end{align}
and defining $\hat{k} \equiv 2\sin\bigl(\frac{k}{2}\bigr)$ and $P_n\bigl(\hat{k}^2\bigr)=P_n\bigl(\bigl(2\sin\bigl(\frac{k}{2}\bigr)\bigr)^2\bigr) = \sin(nk) / \sin\bigl(\frac{k}{2}\bigr)$, leading to
\begin{equation}
 \Pi(\hat{k})-\Pi(0) = i \sum\limits_{n=1}^{N/2-1} \hat{k}\, P_n \,\bigl(\hat{k}^2\bigr)  \overline{\Pi}(n) \,.
\end{equation}
The function $P_n \,\bigl(\hat{k}^2\bigr)$ is related to Chebyshev
polynomials of the second kind and hence analytic in $(-\infty, +1)$,
allowing to evaluate $\Pi(\hat{k})$ at any intermediate
value. Dropping the factor $\hat{k}$ in the above expression, we
obtain the desired expression for the neutron electric dipole form
factor without explicit momentum factors 
\begin{equation}
 \frac{F_3(\hat{k}^2)}{2m_N} = i\sum\limits_{n=1}^{N/2-1} P_n(\hat{k}^2) \, \overline{\Pi}(n) \,,
 \label{eq:method2}
\end{equation}
where we assume that all suppressed kinematic factors have been
included in $\overline{\Pi}(n)$. This expression can be computed
exactly on the lattice -- up to the aforementioned, additional cutoff in the initial Fourier transform -- for any reasonable value of $\hat{k}^2$,
resulting in a smooth curve for $F_3(Q^2)$. Consequently,
taking the statistical errors of the input data into account via
resampling yields a smooth error band for the form factor, as we will
demonstrate in \sec{sec:results_momentum_elimination}. \par 

It is now straightforward to extend the approach to arbitrary sets of
off-axis momentum classes 
\begin{equation}
 M(Q_1,Q_\mathrm{off}^2) = \l\{{\vec{q}} \ | \ {\vec{q}}=\{\pm  Q_1 , Q_2, Q_3 \} \,, \  Q_2^2+Q_3^2=Q_\mathrm{off}^2 \r\}\,,
 \label{eq:off_axis_momenta}
\end{equation}
where $\{\pm Q_1 , Q_2, Q_3\}$ denotes all permutations of $\pm Q_1$, $Q_2$
and $Q_3$. However, to combine the results for $F_3(Q^2)$ for
different $Q_\mathrm{off}^2$--classes as a function of continuous
Euclidean momenta $Q^2=Q^2(\hat{k}, Q_\mathrm{off}^2)$ we need to
consider an analytic continuation for classes with
$Q_\mathrm{off}^2>0$ to reach zero total momentum, i.e. $Q^2=0$. For
our case this amounts to replacing $k\rightarrow i\kappa$ and $\hat{k}
\rightarrow i\hat{\kappa} = -2\sinh\bigl(\frac{\kappa}{2}\bigr)$ in
the derivation outlined above. Note that this also affects $P_n$,
i.e. $P_n\bigl(\hat{\kappa}^2\bigr) = \sinh(n\kappa) /
\sinh\bigl(\frac{\kappa}{2}\bigr)$. In order to obtain the final
result we combine the results from several sets of momentum classes
$M(Q_1,Q_\mathrm{off}^2)$ by taking the error weighted average of the
separate results. \par 

Finally, we remark that in principle Eq.~(\ref{eq:plateau_Pj}) could
also be used instead of (\ref{eq:plateau_P0}) to calculate $F_3(0)$. However, in
this case, not all of the terms share the same momentum prefactor 
and one expects the signal from this
decomposition to be weaker than the one from Eq.~(\ref{eq:plateau_P0})
due to the additional momentum prefactor for $F_3(Q^2)$. Moreover, from a
technical point of view the removal of the double momentum factor $Q_j
Q_k$ is more involved. Therefore, we restrict ourselves to the
combination of current and projection indices given in Eq.~(\ref{eq:plateau_P0}).
\par 

\section{The topological Charge}
\label{sec:The_Topological_Charge}
As already demonstrated in the previous section the evaluation of the
$F_3(Q^2)$ form factor requires the computation of the topological charge, ${\cal
  Q}$ defined in Eq.~(\ref{eq:topological_charge}).
In practice, any valid lattice discretization of $q(x)$ leading to the
right continuum expression of the charge density given in
Eq.~(\ref{eq:Topological_Charge_Density})
could be used for the evaluation of ${\cal Q}$ on the lattice. Here we
choose an improved definition: 
\bea
q(x)=c_0 q_L^{\rm clov}(x) + c_1 q_L^{\rm rect} (x)\, ,
\eea
with 
\bea
q_L^{\rm clov}(x) = \frac{1}{32 \pi^2} \epsilon_{\mu \nu \rho \sigma} {\rm Tr} \left( C^{\rm clov}_{\mu \nu} C^{\rm clov}_{\rho \sigma}  \right) \quad {\rm and} \quad q_L^{\rm rect}(x) = \frac{2}{32 \pi^2} \epsilon_{\mu \nu \rho \sigma} {\rm Tr} \left( C^{\rm rect}_{\mu \nu} C^{\rm rect}_{\rho \sigma}  \right)\, ,
\eea
where
\bea
 C^{\rm clov}_{\mu \nu}(x)= \frac{1}{4} {\rm Im} \left( \parbox{0.8cm}{\rotatebox{0}{\includegraphics[height=0.8cm]{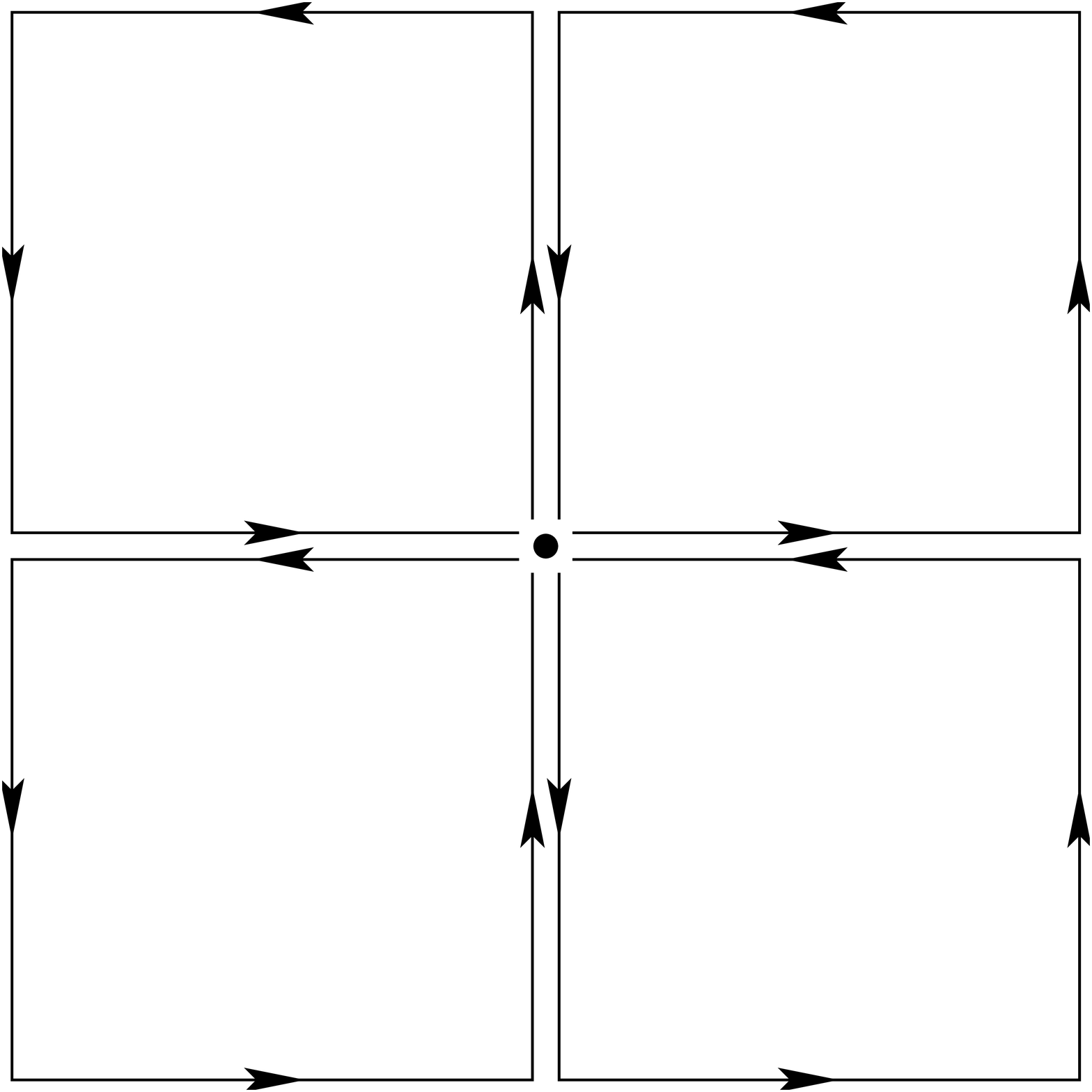}}} \right) \quad {\rm and} \quad  C^{\rm rect}_{\mu \nu}(x) = \frac{1}{8} {\rm Im} \left( \parbox{1.6cm}{\rotatebox{0}{\includegraphics[height=0.8cm]{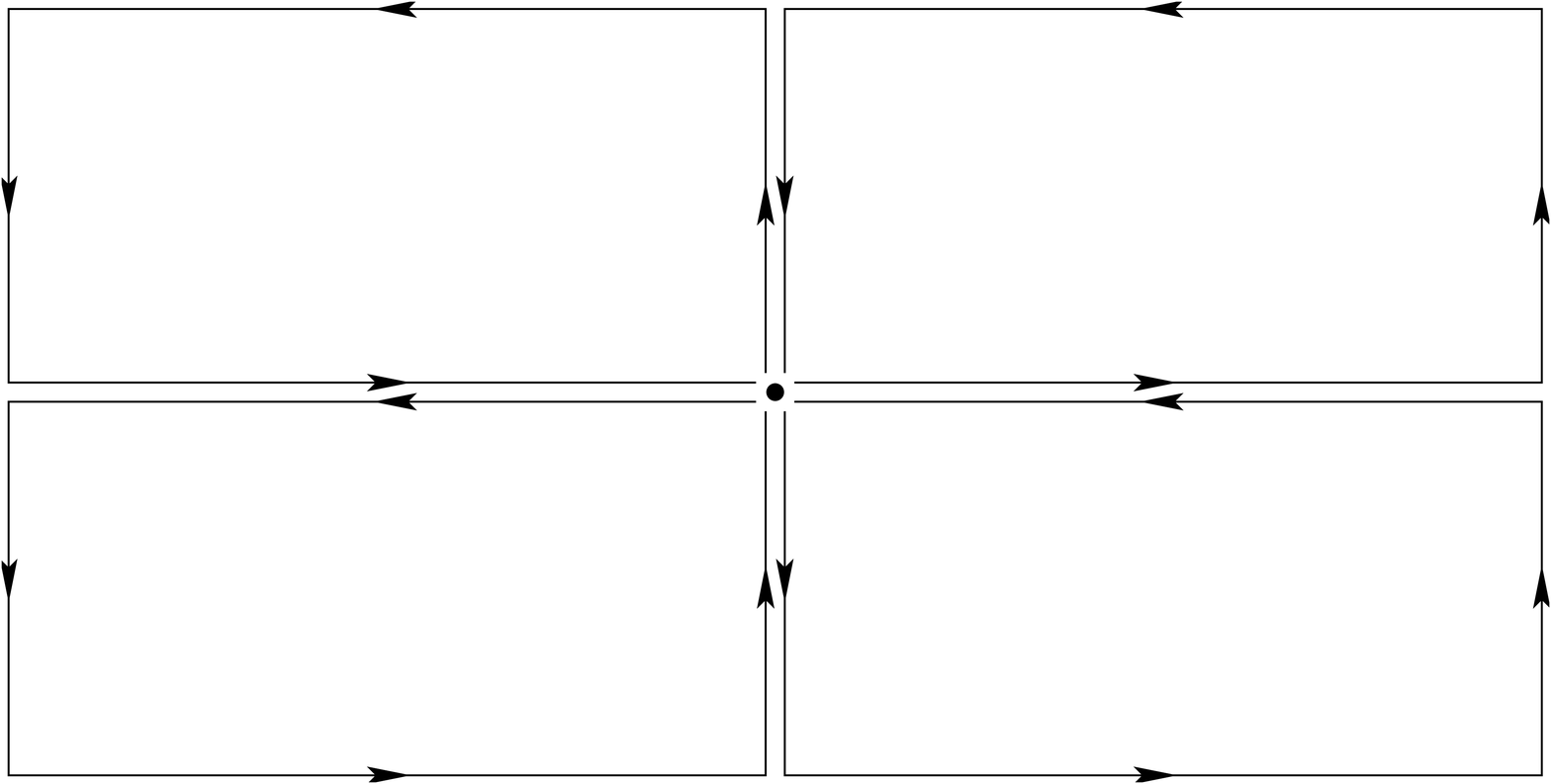}}} + \parbox{0.8cm}{\rotatebox{0}{\includegraphics[height=1.6cm]{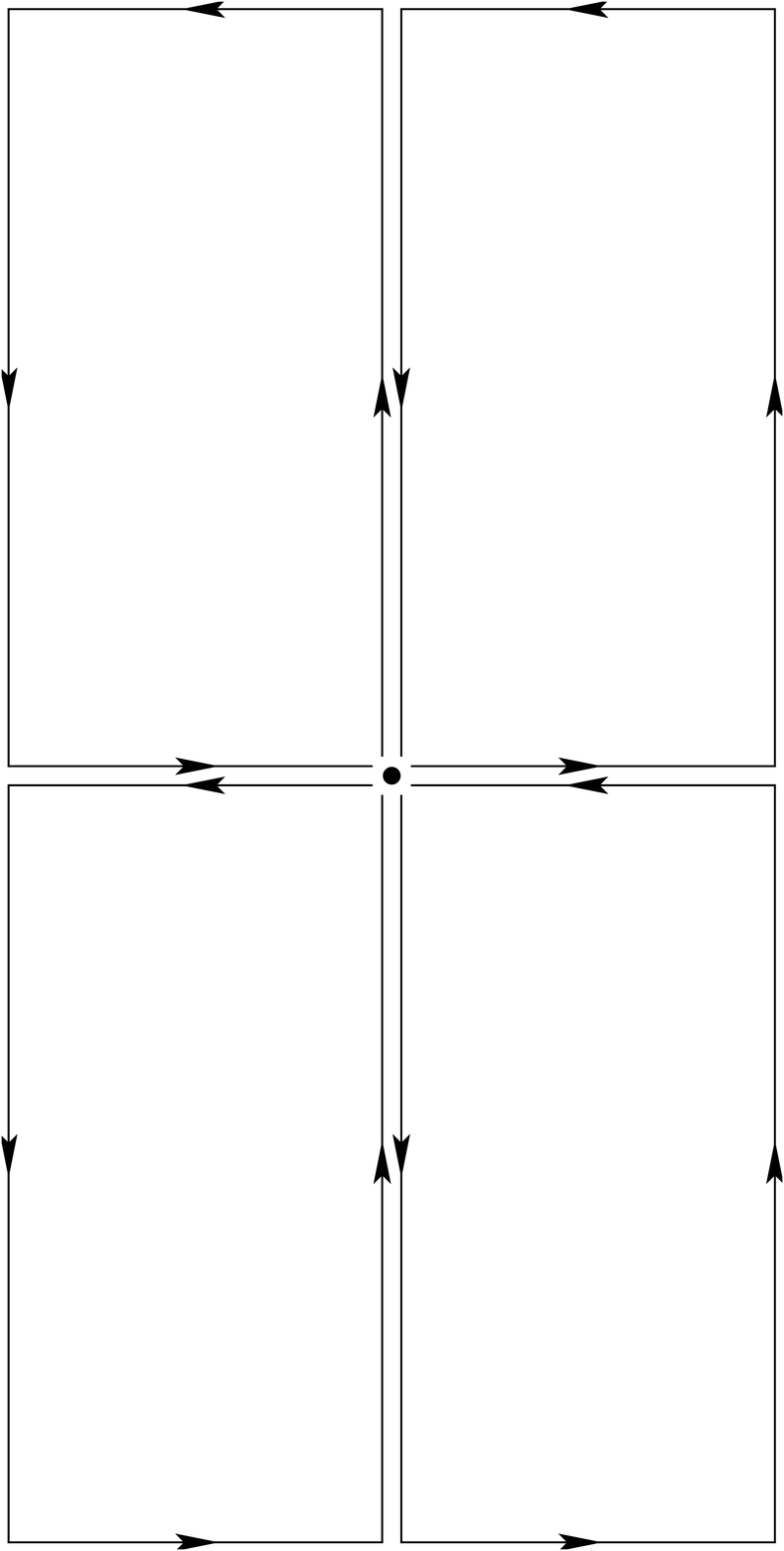}}} \ \right)\,.
\eea
In order to remove the discretization error at tree-level, we use the
coefficients $c_0=5/3$ and $c_1=-1/12$. The lattice operator used for
the evaluation of the topological charge on thermalized configurations
suffers from  ultraviolet fluctuations of the gauge configurations,
hence leading to non-integer results. It is thus customary to use a
smoothing technique, which damps these fluctuations by minimizing the
action locally, without destroying the underlying topological
structure. \par

Such techniques include cooling and the more recently introduced
gradient flow~\cite{Luscher:2009eq,Luscher:2010iy}. It was shown
recently~\cite{Bonati:2014tqa} that both techniques
provide similar results on topological observables such as the average
action and the topological susceptibility when smoothing is done with
the ordinary Wilson action. In Ref.~\cite{Alexandrou:2015yba} we show
that this is also true for actions that include a rectangular term
such as the Iwasaki and the Symanzik tree-level improved action. The
equivalence is realized by the leading order perturbative
rescaling~\cite{Alexandrou:2015yba}
\bea
\tau \simeq \frac{n_c}{3-15 c_1},
\label{eq:perturbative_expression}
\eea
where $\tau$, the gradient flow time, and $n_c$, the number of cooling
steps, are the smoothing scales for gradient flow and cooling, respectively.

We apply both techniques, namely cooling and gradient flow, on the
configurations of the  B55.32 ensemble using the naive Wilson, the
Symanzik tree-level improved and the Iwasaki actions. Regarding the
gradient flow we investigate how the elementary integration step
$\epsilon$ affects our results and find that setting $\epsilon=0.02$
is a safe option; as a matter of fact we observe that smaller
elementary integration steps give indistinguishable results. \par 

While cooling, we measure the improved definition of the
topological charge for every cooling step. Since the gradient
flow is more expensive, we avoid taking measurements for every
integration step, but instead we compute the topological charge every
$\Delta \tau=0.1$; this corresponds to five integration steps. We
cover in total $80-100$ cooling steps while for gradient flow we fix
the maximum gradient flow time according to the perturbative
expression of \eq{eq:perturbative_expression} taking for  $n_c$  the maximum number
of cooling steps. The cooling/gradient flow rescaling factors are
given in \tbl{tab:rescaling}. \par
\begin{table}[ht]
\begin{center}
\begin{tabular}{lccccc}
\hline \hline Smoothing action & $c_0$ & $c_1$ & $n_c/\tau$ & {{$n_c\, (\sqrt{8 t} \approx 0.6 {\rm fm})$}} & {{$\tau\, (n_c)$}} \\
\hline \hline Wilson &  1 & 0 & 3 & 20 & { 6.7}\\
Symanzik tree-level improved  & $\frac{5}{3}$ & $-\frac{1}{12}$ & 4.25 & {{30}} & 7.1 \\
Iwasaki & 3.648 & -0.331  & 7.965 & {{50}} & { 6.3} \\ \hline \hline 
\end{tabular}
\caption{\label{tab:rescaling} The first and second columns give the values
of the parameter $c_0$ and $c_1$ entering in gauge action and the third gives the ratio for $n_c/\tau$  extracted using
  \eq{eq:perturbative_expression}. The fourth column gives  the leading order
  perturbative rescaling between the number of cooling steps and the
  gradient flow time such that the two smoothing techniques are
  equivalent, the fifth column gives  the value of the
  cooling step which corresponds to fixing $\sqrt{8 t} \approx 0.6$~fm,  and the sixth column the
  associated gradient flow time.}
\end{center}
\end{table}

\begin{figure}[htb]
\vspace{-3cm}
\centerline{\hspace{0.0cm}\includegraphics[scale=0.6,angle=270]{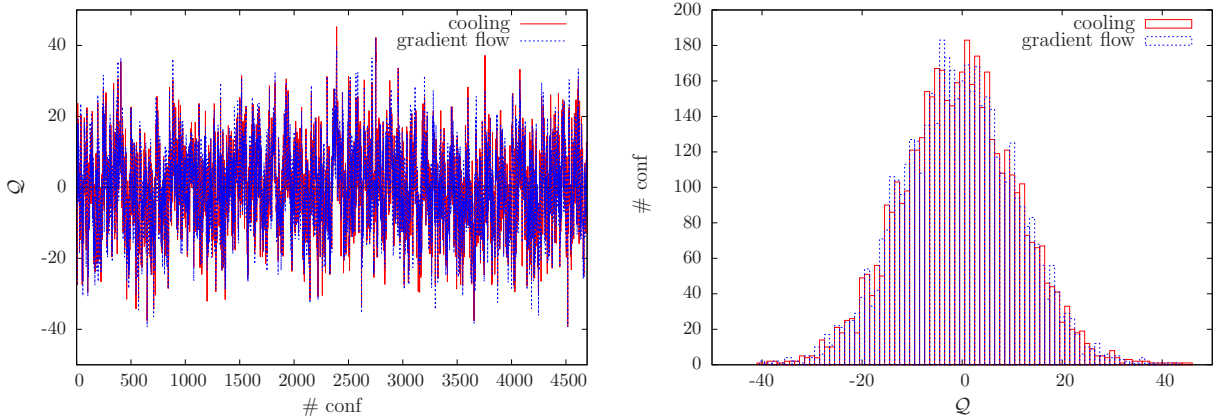}}
\vspace{-3.5cm}
\caption{\label{fig:sampling.tex} The time history of the topological
  charge (left panel) and its associated distribution (right
  panel). The charge has been obtained via cooling (red) and gradient
  flow (blue) with Iwasaki action with black $n_c=50$ and $\tau=6.3$,
  respectively.}
\end{figure}

 According to Ref.~\cite{Luscher:2010iy} one reads 
an observable, whose value depends on the topological charge, at a fixed value
of $\sqrt{8 t}=O(0.1\, {\rm fm})$. The gradient-flow
time $t$ is chosen  such that it is large enough so the relevant observable
has small discretization effects but at the same time small  enough 
 so that the topological content of the fields is
preserved. The observable under consideration here is $F_3/ 2 m_N$
and it  should be scale invariant and, thus, show a plateau  as a function of $\tau$ and $n_c$ at the fixed value of $\sqrt{8 t}$ being within the plateau region. 
For practical reasons we choose
a value of $\tau$ that satisfies $\sqrt{8 t}\approx 0.6 {\rm fm}$ and corresponds to
the rounded value of $n_c$ taken in steps of ten; the associated values of $n_c$ as
well as $\tau$ can be found in \tbl{tab:rescaling}. 
In Ref.~\cite{Alexandrou:2015yba} we demonstrate in more details that 
topological quantities on the same ensemble, for the three smoothing
actions, become equal after a small number of $n_c$ and the equivalent 
gradient flow time; as a matter of fact for $\sqrt{8 t} \approx 0.6
{\rm fm}$ the 
 equality between the smoothing procedures of the gradient flow and cooling is satisfied. In \fig{fig:sampling.tex} we
present the time history of the topological charge ${\cal Q}$ obtained
with cooling using $n_c=50$ and the gradient flow at $\tau=6.3$ 
using the Iwasaki action. This plot demonstrates that the
topological charge does not suffer from large autocorrelations along
the sampling time line with integrated autocorrelation time of $\tau_{\rm int} = 2.6(1)$ and there is a high correlation of $94 \%$ \cite{Alexandrou:2015yba} between
results obtained using cooling and the gradient flow. The latter is a
result of the equivalence between the two
procedures. Additionally, in \fig{fig:sampling.tex},  we provide the
associated histograms of the topological charge for both cooling and
gradient flow, which is found to be approximately Gaussian (according to Anderson-Darling test~\cite{Anderson:1954}). These
observations suggest that for both smoothing approaches the sampling
of the topological charge is adequately good. Similar results are also
obtained for the Wilson and Symanzik tree-level improved
actions. These are the basic requirements that the topological charge
should obey in order to give reliable results for the nEDM.

\section{Results}
\label{sec:Results on nEDM}

In this section, we present our results on the $CP$-odd form factor $\lim_{Q^2\rightarrow 0}F_3(Q^2)$
with our main focus being the extraction of the nEDM,  
using the three approaches discussed in Section~\ref{sec:Plateau_method_to_extract_F_3}. Namely we use a dipole Ansatz to perform an  extrapolation to $Q^2=0$, as well as, the
two position space techniques. For
all three approaches we need the evaluation of $\alpha^1$, which enters
in the determination of $F_3(0)$, as shown in \eq{eq:plateau_F3_comb}. 
Hence, we first discuss the determination of $\alpha^1$.

\subsection{Calculation of $\alpha^1$}
\label{sec:calculation_of_alpha}
To extract $\alpha^1$ we calculate the two-point function 
$G_{{\rm 2pt},{\cal Q}}(\vec{q},t_f,\Gamma_5)$, which 
involves the topological charge ${\cal Q}$, as well as the usual
two-point function $G_{\rm 2pt}(\vec{q},t_f,\Gamma_0)$. Note that the
argument $t_i$ has been omitted, since in our calculation we shift the
source point to $t_i=0$. We form the ratio of these two-point
functions according to \eq{eq:2_pt_function} at zero momentum transfer,
$\vec{q}=\vec{0}$. The topological charge is computed using both
cooling and the gradient flow method employing the  three gauge actions,
  Wilson, Symanzik tree-level improved and Iwasaki. 
For the calculation of the $\alpha^1$ parameter we use a large
statistical sample of a total of 36720 two-point functions (2295 configurations, each
with 16 source positions). This allows us to decrease significantly
the statistical errors on $\alpha^1$. As a consequence, the bins on
which $\alpha^1$ is computed, do not coincide with those of $F_3$, and
thus we need an alternative procedure to jack-knife for the computation of the
statistical errors on the nEDM. This will be
discussed in Subsection~\ref{sec:Calculation_of_F3}.

\begin{figure}[h!]
\vspace{-1cm}
\centerline{\hspace{0.0cm}\includegraphics[scale=0.4,angle=270]{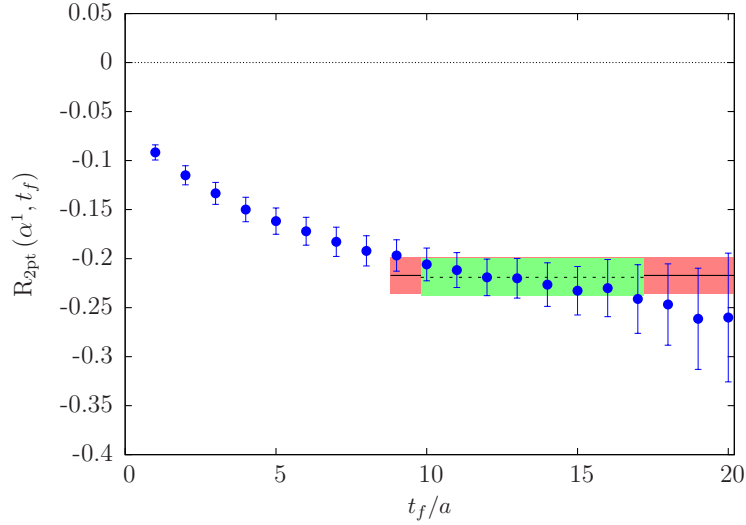}}
\caption{\label{fig:2pt_plateau_fig} 
The ratio ${\rm R}_{\rm 2pt}(\alpha^1,t_f)$ as a function of
$t_f/a$ ($t_i$ was set to 0). A constant fit between the
time-slices 9-20 gives the solid line $\alpha^1=-0.217(18)$ with $\chi^2 / {\rm DOF}=0.51$ and
the red band is the associated statistical uncertainty. The green
band corresponds to the fit range 10-17 with $\chi^2 / {\rm DOF}=0.29$, which is fully compatible with
the red one. The employed topological
charge ${\cal Q}$ is extracted via the gradient flow at $\tau=6.3$
using the Iwasaki action.}
\end{figure}

The ratio ${\rm R}_{\rm 2pt}(\alpha^1,t_f)$ is shown in
\fig{fig:2pt_plateau_fig} as a function of $t_f/a$, for which the
topological charge is measured using the gradient flow with the
Iwasaki action at  $\tau=6.3$. As can be seen, for $t_f/a>8$ the ratio
becomes time-independent yielding $\Pi_{\rm 2pt} (\alpha^1)$. By
fitting to a constant in the plateau region within the time interval 9-20 we extract
the value of $\alpha^1=-0.217(18)$.  We have checked that by modifying both the
starting and ending time-slices in the fit within the region 
$t_f/a=7$ to 22 we get compatible results. An example is shown in
\fig{fig:2pt_plateau_fig} for the range $t_f/a=10$ to 17, which gives
the green band. 

\begin{figure}[h!]
\vspace{-3cm}
\centerline{\hspace{0.0cm}\includegraphics[scale=0.625,angle=270]{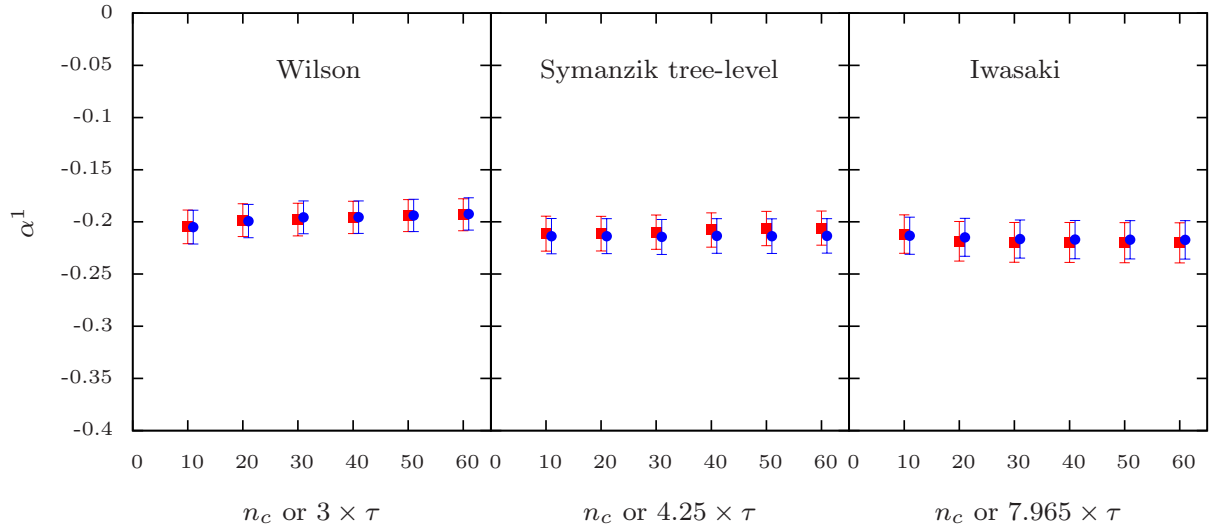}}
\vspace{-3cm}
\caption{\label{fig:alpha_vs_nc} The value of $\alpha^1$ as a function
  of $n_c$ or $3\times \tau$ (left), $4.25\times \tau$ (center) and
  $7.965\times\tau$ (right) for Wilson, Symanzik tree-level improved
  and Iwasaki smoothing actions, respectively. Data obtained by cooling
  are shown by the red squares, while data extracted via the
  gradient-flow are shown by the blue circles and have been
  shifted horizontally to be more visible.} 
\end{figure}

The final result on $\alpha^1$ should not depend on the smoothing
scale; in other words there should be a plateau for some region of the
smoothing scale $n_c$ or equivalently $\tau$. This is demonstrated in
\fig{fig:alpha_vs_nc} where we  plot the value of $\alpha^1$ as a
function of $n_c$ and the gradient flow time $\tau$ rescaled according
to \eq{eq:perturbative_expression} for all three gauge actions. As can
be seen, the $\alpha^1$ parameter remains unchanged after  $n_c \geq 20$ cooling
steps or the equivalent gradient flow time
$\tau$. This is in perfect agreement with fixing $\sqrt{8 t} \approx 0.6 {\rm fm}$. Therefore, setting $n_c$ according to the fifth column in \tbl{tab:rescaling} or fixing to the corresponding flow time for the gradient flow both yield a consistent value for
$\alpha^1$. We will employ the value $\alpha^1=-0.217(18)$, obtained
using the gradient flow at $\tau=6.3$ for the Iwasaki action in order
to determine $F_3(Q^2)$. We also observe that for all gauge actions the value of
$\alpha^1$ extracted when using cooling or the gradient flow to
determine the topological charge are in agreement, reflecting the
equivalence between the two procedures. The results shown in
Fig.~\ref{fig:alpha_vs_nc} use the topological charge obtained for
every $\Delta n_c=10$ starting from $n_c=10$ while cooling and for the
corresponding gradient flow time $\tau$ when smoothing via the gradient
flow. The same computation of the topological charge is also used in
the evaluation of the three-point function. For the determination of
the errors we use jack-knife with bin size of $5$; larger bin sizes
give consistent results. \par

\subsection{Results for $F_3(0)$}
\label{sec:Calculation_of_F3}
\subsubsection{$F_3(0)$ via extrapolation in $Q^2$}
\label{sec:results_for_dipole_fit}
We first discuss the determination of $F_3(0)$ extracted by fitting
the $Q^2$-dependence of $F_3(Q^2)$ to the dipole Ansatz given in
\eq{eq:dipole_fit}. $F_3(Q^2)$ has been computed for a sequence of
values of the momentum transfers, $Q^2=2m_N \sqrt{E_N - m_N}$, with the 
momentum chosen such that the spatial components $Q_i$ take all
possible combinations of $Q_i / (2\pi/L) \in [0,\pm 4]$ (and all
permutations thereof). We perform the calculation at
three values of the source-sink separation namely
$t_{\rm sep}=10a$,  $t_{\rm sep}=12a$ and $t_{\rm sep}=14a$ with  
 the statistics being 2357 for $t_{\rm sep}=10a$, 
 and 4623  for $t_{\rm sep}=12a, 14a$. 
\begin{figure}[h!]
\vspace{-1cm}
\centerline{\hspace{0.0cm}\includegraphics[scale=0.45,angle=270]{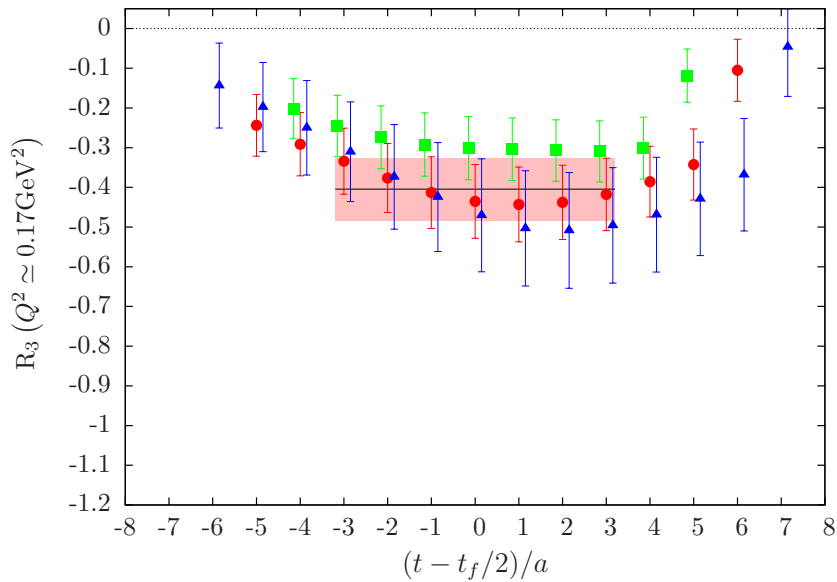}}
\vspace{-0.5cm}
\caption{\label{fig:3pt_plateau} The ratio leading to $F_3$
  (Eq.~(\ref{eq:plateau_F3_comb})) for the first non-zero momentum transfer
  ($Q^2 \simeq 0.17\, {\rm GeV}^2$) as a function of the insertion time
  $(t-t_f/2)/a$. Green squares, red circles and blue triangles correspond
  to source-sink separation of $t_f/a=10,\,12,\,14$, respectively. The
  topological charge is evaluated using the gradient flow at
  $\tau=6.3$ with the Iwasaki action. }
\end{figure}

In \fig{fig:3pt_plateau} we show the results for the combination of
ratios leading to the extraction of $F_3(Q^2)$ according to 
\eq{eq:plateau_F3_comb}. The ratio of $F_3$ is plotted for three source-sink time
separations: $10a,\,12a,\,14a$ corresponding to $0.82,\,1.0,\,1.15$ fm,
respectively. As can be seen the results at the three sink-source time
separations are consistent. The ratio shown
corresponds to the lowest non-zero momentum transfer, that is
$Q^2\simeq0.17\, {\rm GeV}^2$, upon averaging over the data for the
two spatial components $Q_i/(2\pi/L)=(\pm 1,0,0)$ and three
permutations. In what follows  we  will use $t_{\rm sep}=12a$, which yields a better
statistical accuracy and it is fully consistent with the
results obtained for $t_{\rm sep}=14a$. For $t_{\rm sep}=12a$, a plateau can be identified
in the range $t/a \in [3,9]$, from which we extract $F_3(Q^2\simeq0.17 {\rm GeV}^2)$ 
(solid line in \fig{fig:3pt_plateau}).  %
 However, for a proper computation of the error we cannot use jackknife since, as  mentioned in Section~\ref{sec:calculation_of_alpha}, 
the $\alpha^1$ parameter is computed on 2295 configurations with
multiple source positions, while  $F_3$ is computed using  4623 configurations with a single source position,
which does not allow to combine the bin
values of $\alpha^1$ and $F_3$. We thus use the following procedure to 
take into account the statistical error of  
$\alpha^1$  in the evaluation of the error on
$F_3$: We first compute $F_3$ and the
associated jackknife error, $d F_3$ by employing the mean value for
$\alpha^1$. 
 We then recompute $F_3$ using 
 $\alpha^1_{\rm max}=\alpha^1 +
d\alpha^1$, where $d\alpha^1$ is the jackknife error of $\alpha^1$.
We denoted the difference in the values obtained using the mean value of $\alpha^1$ and $\alpha^1_{\rm max}$ by $\Delta F_3$. The final
error on $F_3$  is computed by combining $\Delta F_3$ due to the variation in $\alpha^1$  with the
jackknife error $dF_3$ in quadrature, namely
$\sqrt{(\Delta F_3)^2 + (d F_3)^2}$.  In \fig{fig:3pt_plateau} we show the final error on $F_3$ with a red band. This procedure of
taking into account the error on the $\alpha^1$ parameter is 
 compatible with the  error using a resampling procedure. For the
latter we first generate samples for $\alpha^1$ that are gaussian
distributed with $\sqrt{N_s-1}\, d\alpha^1$, where $N_s$ denotes
the number of jackknife samples in our analysis. These samples are then used in
our jackknife analysis together with the actual jackknife samples for
the remaining quantities.

Once again, we stress that no separate calculation of
$F_1(Q^2)$ and $F_2(Q^2)$ is needed, since $F_3(Q^2)$ is extracted
from a combination of ratios leading to \eq{eq:plateau_F3_comb}, upon
substituting the expressions for $G_E(Q^2)$ and $G_M(Q^2)$
(\eq{eq:q_G_E} and \eq{eq:q_G_M}) at the level of the time-dependent
ratios. 
In addition, the value of the nucleon mass, $m_N$, enters in
\eq{eq:plateau_F3_comb}, which is calculated using the nucleon two-point
function on the same configurations analyzed for $F_3(Q^2)$ and in the
same bin. The value of $am_N$ in lattice units is presented
in~\tbl{Table:params}. \par

\begin{figure}[h!]
\vspace{-1cm}
\centerline{\hspace{0.0cm}\includegraphics[scale=0.45,angle=270]{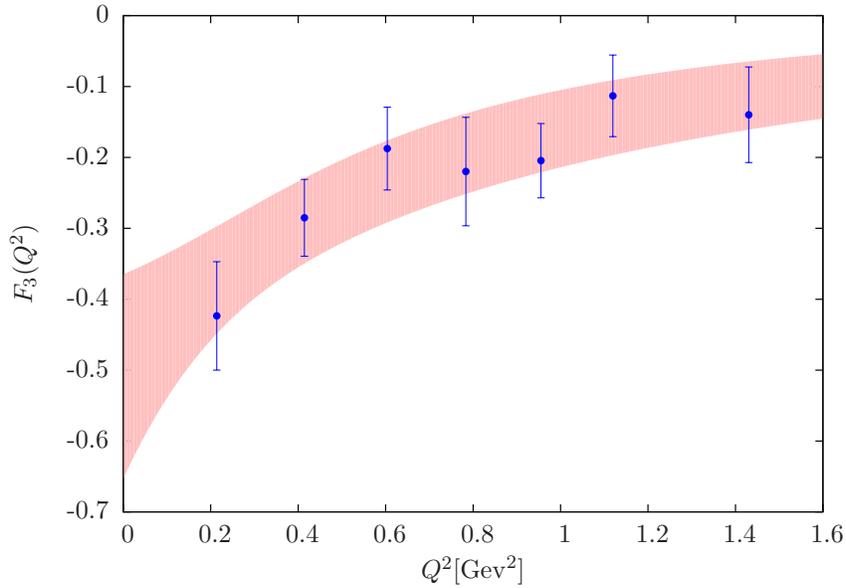}}
\vspace{-0.5cm}
\caption{\label{fig:dipole_fit} $F_3(Q^2)$ versus $Q^2$ for the same
parameters as in  Fig.~\ref{fig:3pt_plateau} for $t_{\rm
  sep}=12a$. The band is the resulting dipole fit using the form given in
\eq{eq:dipole_fit} to $F_3(Q^2)$ and data for $Q^2<1$~GeV$^2$. The fit
gives $F_3(0) =-0.509(144)$, $am_{F_3}=0.469(133)$ with $\chi^2 / {\rm DOF}=0.54$.}
\end{figure}

After determining $F_3(Q^2)$ by identifying the corresponding plateau
at each value of $Q^2$ we perform a fit using the  dipole form of \eq{eq:dipole_fit}, treating $F_3(0)$ as a fitting parameter.
In
Fig.~\ref{fig:dipole_fit} we show the resulting fit when  the Iwasaki smoothing action and
gradient flow with $\tau=6.3$ are used in the calculation of the topological charge.  The fit is performed  for $Q^2<
1\,{\rm GeV}^2$. To check for fit stability we vary the upper
limit of the fit range, $Q^2_{\rm max}$, from 0.8 to 1.5 ${\rm
GeV^2}$, and found consistent results for $F_3(0)$. 

\begin{figure}[h!]
\vspace{-3cm}
\centerline{\hspace{0.0cm}\includegraphics[scale=0.625,angle=270]{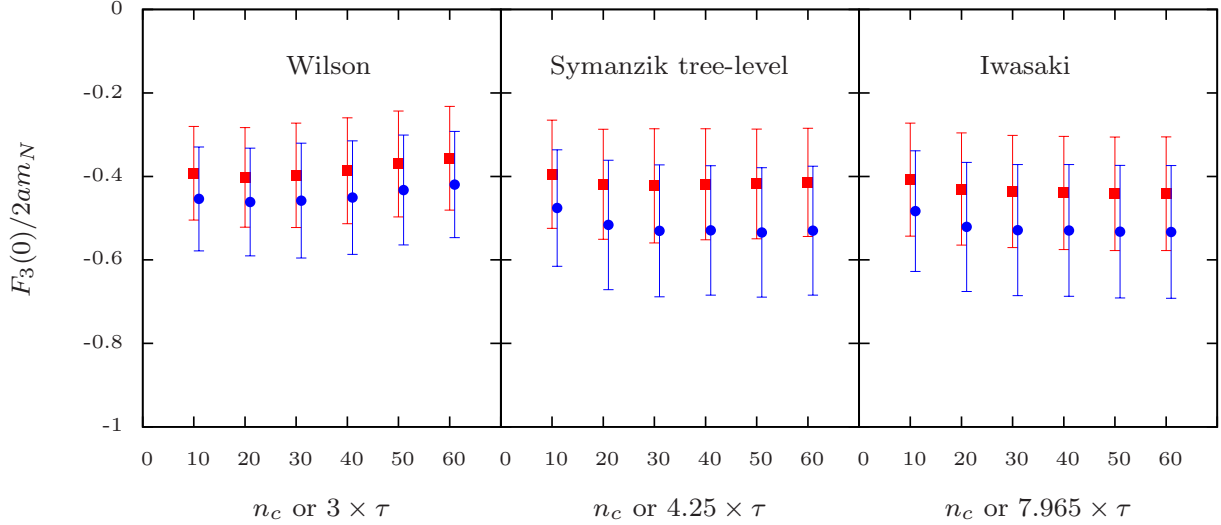}}
\vspace{-3cm}
\caption{\label{fig:nedm_vs_nc}  The nEDM in lattice units as a function of
  $n_c$ or $3\times \tau$ (left), 
  $4.25\times \tau$ (center) and $7.965\times\tau$ (right) for Wilson,
  Symanzik tree-level improved and Iwasaki gauge actions,
  respectively. The results are obtained using a fit to the dipole form of
  \eq{eq:dipole_fit} with a cutoff $Q^2 < \left( 1 \ {\rm GeV}
  \right)^2$. Data obtained by cooling (gradient-flow) are shown by
  red squares (blue circles). Gradient-flow results have been shifted
  horizontally to improve visibility.}
\end{figure}
\FloatBarrier
 In \fig{fig:nedm_vs_nc} we present $F_3(0)/(2 a m_N)$ as a function of
the number of cooling steps $n_c$ and the corresponding gradient flow
time.  These results clearly confirm that the value of the nEDM obtained
via cooling agrees with the one extracted via the gradient flow
corroborating  the equivalence of cooling and gradient flow also on
the level of the nEDM. In addition, we observe that $\vert \vec{d}_N
\vert / {\bar \theta}$ (\eq{eq:dN}) is adequately stable for $n_c \geq 20$ and the
corresponding gradient flow time. By looking at  the nEDM as a function of $\tau$ and $n_c$ for the gradient flow and cooling respectively we observe that fixing  $\sqrt{8 t} \approx 0.6 {\rm
  fm}$ is in the plateau region and suggests that the value
of $F_3(0)/(2 m_N)$ can be taken for $n_c = 20, \ 30 \ {\rm and} \ 50$  cooling  steps or at the corresponding gradient flow times for Wilson, Symanzik tree-level improved and Iwasaki actions, respectively. The results for $F_3(0)/(2m_N)$ extracted using the dipole fit and  the above cooling steps or the corresponding gradient flow times are collected in \tbl{tab:nedm_dipole_fit}. As can be seen,
the values obtained for $t_{\rm sep}=14a$ are consistent with those obtained
using $t_{\rm sep}=12a$ albeit with larger errors. 
We use data obtained for $t_{\rm sep}=12a$ to determine our final result due to their better statistical accuracy compared to the results obtained for $t_{\rm sep}=14a$. In addition we stress that we do not attempt to make an extrapolation to infinite volume i.e $t_{\rm sep} \to \infty$ and determine the corresponding systematic error.
\begin{table}[h!]
 \centering
 \begin{tabular}{@{\extracolsep{\fill}}ccccc}
\hline
\multicolumn{5}{c}{$F_3(0)/(2m_N)$ (e.fm)}\\
\hline\hline
Gauge action & \multicolumn{2}{c}{cooling} & \multicolumn{2}{c}{gradient flow} \\
             & $t_{\rm sep}=12a$ & $t_{\rm sep}=14a$ & $t_{\rm sep}=12a$ & $t_{\rm sep}=14a$ \\ 
\hline\hline
Wilson    &-0.035(09)  &-0.056(45)  &-0.039(10)  &-0.065(42)  \\
Symanzik  &-0.036(10)  &-0.072(50)  &-0.046(12)  &-0.076(40)  \\
Iwasaki   &-0.035(10)  &-0.049(22)  &-0.041(12)  &-0.053(23)  \\
  \hline\hline
 \end{tabular}
\caption{\label{tab:nedm_dipole_fit}$F_3(0)/(2m_N)$ extracted from fitting to
  the dipole Ansatz of \eq{eq:dipole_fit}. We include results for both $t_{\rm sep}=12a$ and $t_{\rm sep}=14a$ at $n_c=20$, $n_c=30$ and $n_c=50$ as well as at the corresponding gradient
  flow times given in Table~\ref{tab:rescaling} for the Wilson, Symanzik tree-level improved and Iwasaki actions, respectively.
}
\end{table}

\subsubsection{$F_3(0)$ via the application of the derivative to the ratio technique}
In this section we discuss the results on $F_3(0)$ using the
application of the derivative to the ratio method. As explained  in \sec{sec:continuum_derivative_technique}, this requires the 
construction of the derivative-like three point functions expressed
as $\sum_{x_j = -L/2 + a}^{L/2 - a} ( \sum_{\substack{x_i=0, i \ne j}}^{L -a} i x_j G^\mu_{{\rm 3pt}, {\cal Q}} (\vec{x}, t_f, t_i, t, \Gamma_k))$,
$\sum_{x_j = -L/2 + a}^{L/2 - a} ( \sum_{\substack{x_i=0, i \ne j}}^{L -a} i x_j G^k_{{\rm 3pt}} (\vec{x}, t_f, t_i, t, \Gamma_0))$
and $\sum_{x_j = -L/2 + a}^{L/2 - a} ( \sum_{\substack{x_i=0, i \ne j}}^{L -a} i x_j G^k_{{\rm 3pt}} (\vec{x}, t_f, t_i, t, \Gamma_i))$
as well as the two-point function $G_{\rm 2pt}(\vec{q},t_f,0,\Gamma_0)$ in order to form the right ratios. The three derivative-like three point functions are computed  by taking the  Fourier
transformation according to \eq{eq:summation}. As previously
mentioned, we use a source-sink separation of $t_{\rm sep}=12a$ and shift
the source timeslice to $t_i=0$. Once more we check for ground state dominance by extracting the nEDM for $t_{\rm sep}=10a$ and $t_{\rm sep}=14a$. In addition, we average over the
spatial direction $j=1,2,3$ as this appears in~\eq{eq:Continuum_derivative_plateau}. 
We fit the ratio to a constant in the plateau region to extract the quantity 
given in \eq{eq:Continuum_derivative_plateau}  and use the procedure explained in \sec{sec:results_for_dipole_fit} by employing the mean value for $\alpha^1$ as well as $\alpha^1_{\rm max}=\alpha^1 + d\alpha^1$ in order to compute the associated statistical error on $F_3$. As discussed in~\sec{sec:continuum_derivative_technique} there can be a residual time dependence  in \eq{eq:Continuum_derivative_plateau} of the form $\sim {\rm exp} \left(  a (E_N (\vec{q}) -m_N) \ t/a  \right)$. Hence, in addition to the constant fit  we also perform an exponential fit in order  to provide a systematic error. We take  the difference between the value determined from  the constant fit and that extracted when we include the 
exponential time-dependence as the systematic error. We note that the resulting systematic error is comparable to our statistical error making this approach useful.

\begin{figure}[h!]
\vspace{-3cm}
\centerline{\hspace{0.0cm}\includegraphics[scale=0.625,angle=270]{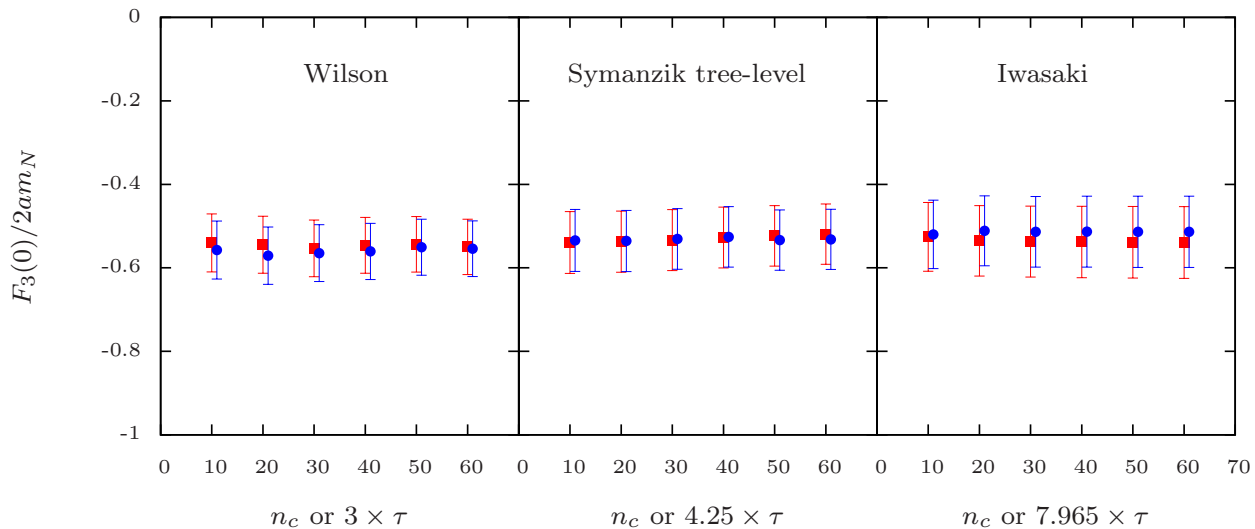}}
\vspace{-3cm}
\caption{\label{fig:nedm_vs_flow_derivative} Results for the nEDM in lattice units using the application of the derivative to the ratio as
  a function of $n_c$ or $3\times \tau$ (left), $4.25\times \tau$ (center) and
  $7.965\times\tau$ (right). The notation is the same as in
  Fig.~\ref{fig:nedm_vs_nc}.  }
\end{figure}

In \fig{fig:nedm_vs_flow_derivative} we show the results for $F_3(0)/(2
a m_N)$ in lattice units. The derivative-like three-point function $\sum_{x_j = -L/2 + a}^{L/2 - a} ( \sum_{\substack{x_i=0, i \ne j}}^{L -a} i x_j G^\mu_{{\rm 3pt}, {\cal Q}} (\vec{x}, t_f, t_i, t, \Gamma_k))$ uses
the topological charge extracted from cooling and the gradient flow for the
Wilson, Symanzik tree-level improved and Iwasaki smoothing
actions as a function of the cooling steps or the flow time $\tau$. 
Similarly to Fig.~\ref{fig:nedm_vs_nc}, the value of the nEDM is stable for
the smoothing scales used to define the topological charge.  Namely, this justifies our choice of fixing $\sqrt{8 t} \approx 0.6{\rm fm}$,  thus results in using $n_c=20$, 30 and 50 as well as to the corresponding gradient flow times for Wilson, Symanzik tree-level improved and Iwasaki smoothing actions respectively. \par

\begin{figure}[h!]
\vspace{-1cm}
\centerline{\hspace{0.0cm}\includegraphics[scale=0.45,angle=270]{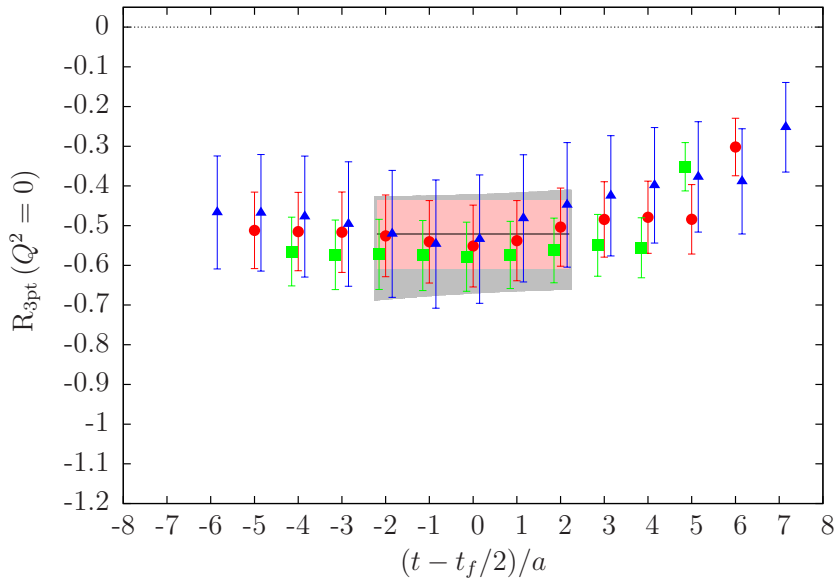}}
\vspace{-0.5cm}
\caption{\label{fig:3pt_plateau_continuum_derivative_method} 
The ratio leading to $F_3(0)$ determined using the continuum
derivative method. Green squares, red circles and blue triangles correspond
to source-sink separation of $t_f/a=10,\,12,\,14$, respectively. The
topological charge is evaluated using the gradient flow at
$\tau=6.3$ with the Iwasaki action. The behaviour when using other smoothing actions to calculate the topological charge is similar.}
\end{figure}

 In \fig{fig:3pt_plateau_continuum_derivative_method} we
  show results for the combination of continuum-like derivatives of
  ratios leading to the extraction of $F_3(0)$ according to 
\eq{eq:Continuum_derivative_plateau}. The results are produced with
topological charge extracted using the gradient flow for a total of
$\tau =6.3$ with the Iwasaki action. A plateau can be identified and
fitted in the range $t/a \in [4,8]$ yielding a value of
$F_3(0)=-0.52(09)$. The statistical
error is computed using the method explained in the previous subsection and is presented by the red band. In addition we provide the exponential fit with its statistical uncertainty  resulting from the residual exponential time-dependence in the three-point function. The fit is done in the range $t/a \in [4,8]$ yielding a value of $F_3(0)=-0.55(13)$. As can be seen, the two bands  yield consistent results.

 In \tbl{tab:nedm_continuum_derivative} we give the results for
$F_3(0)/(2m_N)$  for the three gauge smoothing actions using our standard 
parameters for $n_c$ and $\tau$.  We provide the statistical as well as the systematic
  error induced by the residual exponential time-dependence
in the three-point function (see \eq{eq:summation}).\par
\begin{table}[h]
 \centering
 \begin{tabular}{@{\extracolsep{\fill}}ccccc}
\hline
\multicolumn{5}{c}{$F_3(0)/(2m_N)$ (e.fm)}\\
 \hline\hline
Gauge action & \multicolumn{2}{c}{cooling} & \multicolumn{2}{c}{gradient flow} \\
             & $t_{\rm sep}=12a$ & $t_{\rm sep}=14a$ & $t_{\rm sep}=12a$ & $t_{\rm sep}=14a$ \\ 
\hline\hline
        Wilson    &-0.044(6)(6) & -0.044(8)(8) & -0.046(6)(9) & -0.046(8)(15) \\
	Symanzik & -0.043(6)(3) & -0.043(9)(6) & -0.043(6)(4) & -0.042(9)(10) \\
	Iwasaki   & -0.044(7)(4) & -0.043(11)(10) & -0.042(7)(3) & -0.040(11)(8) \\
  \hline\hline
 \end{tabular}
\caption{\label{tab:nedm_continuum_derivative} 
 Results for $F_3(0)/(2 m_N)$ extracted using the 
{application of the derivative to the ratio} method for $t_{\rm sep}=12a, 14a$. Results are shown for
$n_c=20, \ 30, \ {\rm and} \ 50$ as well as for the corresponding
gradient flow times according to \tbl{tab:rescaling} for Wilson,
Symanzik tree-level improved and Iwasaki actions respectively. The error in the
first parenthesis is statistical and in the second the systematic determined as discussed in the text.}
\end{table}

\subsubsection{$F_3(0)$ with the elimination of the momentum in the plateau region technique}
\label{sec:results_momentum_elimination}
The elimination of the momentum in the plateau region method is an improved technique to  remove
momentum prefactors in the form factor decomposition of a given  ratio
of correlators on the lattice. The details of this method can be found
in Section~\ref{sec:y_summation} and Ref.~\cite{Alexandrou:2014exa}
where the method was applied for the first time for the evaluation of
the nucleon anomalous magnetic moment. Similar approaches using
analytic continuation have been used in the context of calculating
hadronic vacuum polarizations~\cite{Bernecker:2011gh,Feng:2013xsa}. In
the following we present the first results for the nEDM obtained within this approach. \par

\begin{figure}[h!]
\vspace{-5cm}
\centerline{\hspace{0.0cm}\includegraphics[scale=0.775,angle=270]{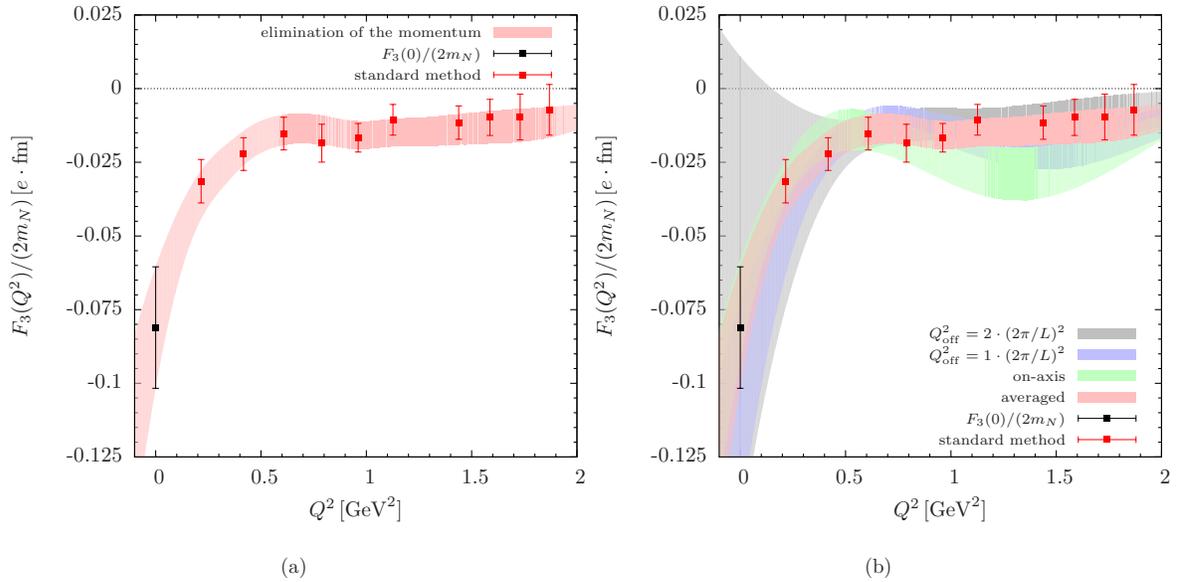}}
\vspace{-3.5cm}
 \caption{\label{fig:ysummation} 
   Results on nEDM using the momentum
   elimination method for source-sink separation
   $t_\mathrm{sep}=12a$. The right panel shows separate bands for the three
   sets of lowest off-axis momentum classes $M({{ Q_1}},Q_\mathrm{off}^2)$.}
\end{figure}

As in the other methods, here we also extract $F_3(0)$ using the
various definitions for the topological charge (Wilson, Symanzik
tree-level improved and Iwasaki actions for both cooling and gradient
flow). The results obtained within this approach 
exhibit a similar behavior as that depicted in~\fig{fig:nedm_vs_nc} and
\fig{fig:nedm_vs_flow_derivative}. We analyze data for two source-sink
separations, namely  of $t_{\rm sep}=12a$ and $t_{\rm sep}=14a$, employing a general momentum
cutoff ${Q_1}<4\cdot(2\pi/L)$ and momentum classes with an off-axis
momentum squared of up to $Q^2_\mathrm{off} \leq 5 \cdot (2\pi/L)^2$
(c.f. Eq.~(\ref{eq:off_axis_momenta})). 
The red bands in both panels of Fig.~\ref{fig:ysummation}
show the results for $F_3(Q^2)/(2m_N)$ extracted using the momentum
elimination method and $t_{\rm sep}=12a$. It is obtained as the error
weighted average over all sets of different off-axis momentum classes
$M(Q_1,Q_\mathrm{off}^2)$. As required, the band reproduces the red
points which are the results obtained using plateau method at each
$Q^2$ value. \par 

The right panel of Fig.~\ref{fig:ysummation} contains separate bands
for the results from different value of $Q_\mathrm{off}^2$, which
agree within their errors. However, the errors at small values of
$Q^2$ grow rapidly with increasing $Q^2_\mathrm{off}$, such that
higher off-axis momentum classes with $Q_\mathrm{off}^2\geq 2 \cdot
(2\pi/L)^2$ hardly contribute to the nEDM at all. \par 

The results obtained using this approach are collected in
Table~\ref{tab:ysummation}. Although using
the Wilson smoothing action to calculate the topological charge 
 (either through cooling or the gradient flow) we obtained 
  lower mean values
and equivalently higher when using the Iwasaki action, the results are compatible within errors. Comparing the
results for $t_{\rm sep}=12a$ and $t_{\rm sep}=14a$ we can not detect
excited state contamination within statistical errors. 

In addition, results obtained using the elimination of the momentum in the plateau region technique, which does not assume
any functional form for the momentum dependence  are compatible with the dipole fit results. The dipole form fit yields correlated $\chi^2 / {\rm DOF}=0.54$,  supporting a dipole behaviour, which is the approach used in other lattice studies.

\begin{table}[h!]
 \centering
 \begin{tabular}{@{\extracolsep{\fill}}ccccc}
\hline
\multicolumn{5}{c}{$F_3(0)/(2m_N)$ (e.fm)}\\
\hline\hline
Gauge action & \multicolumn{2}{c}{cooling} & \multicolumn{2}{c}{gradient flow} \\
             & $t_{\rm sep}=12a$ & $t_{\rm sep}=14a$ & $t_{\rm sep}=12a$ & $t_{\rm sep}=14a$ \\ 
\hline\hline
Wilson    & -0.052(17) & -0.043(26) & -0.056(17) & -0.043(26) \\
Symanzik  & -0.066(18) & -0.053(28) & -0.068(17) & -0.055(28) \\
Iwasaki   & -0.082(21) & -0.076(32) & -0.082(21) & -0.073(32) \\
  \hline\hline
 \end{tabular}
 \caption{Results for the neutron electric dipole moment in physical
   units for two values of the source-sink separation and the three
   different smoothing actions.  Results are given for $n_c$ and $\tau$ which correspond to $\sqrt{8 t}\approx 0.6 {\rm fm}$ according to \tbl{tab:rescaling}.}
 \label{tab:ysummation}
\end{table}

\section{Conclusions}
\label{sec:Conclusions}

\begin{figure}[htb]
\vspace{-1.5cm}
\centerline{\hspace{0.0cm}\includegraphics[scale=0.45,angle=270]{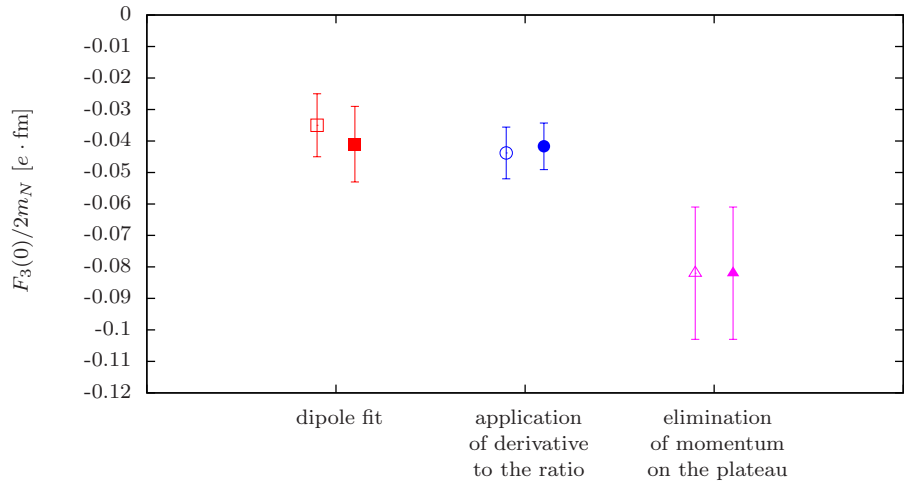}}
\vspace{-1.5cm}
\caption{\label{fig:Final_Result_Iwasaki} Our results for $F_3(0)/(2m_N)$ in physical
units using different approaches to determine $F_3(0)$.  Open (filled) symbols show results
using cooling (gradient flow) for the extraction of the topological charge. Red/blue/magenta points
show results extracted using a dipole fit/{application of the derivative to the ratio}/{elimination of the momentum in the plateau region} approach, respectively. The
topological charge has been evaluated using the Iwasaki smoothing action.}
\end{figure}

The neutron electric dipole moment is computed using $N_f{=}2{+}1{+}1$
twisted mass fermions simulated at a pion mass of 373~MeV and lattice
spacing of $a=0.082$~fm employing a total of 4623 measurements. This high
statistics analysis enables us to extract reliable results on the
$CP$-violating form factor $F_3(0)$ and benchmark our techniques. Due
to the multiplicative kinematical factors appearing in front of $F_3(Q^2)$
it  cannot be extracted
directly from the matrix element. The usual
approach is to extrapolate $F(Q^2)$ to $Q^2=0$ by fitting its
$Q^2$-dependence employing an Ansatz for its momentum dependence. In
this work, besides this standard approach, we employ two new
techniques that explicitly eliminate the kinematical factor yielding
directly $F_3(0)$ without any model assumption on its
$Q^2$-dependence. These techniques involve the three-point function in
coordinate space and two different ways to eliminate the kinematical
factor.  The behaviour of the nEDM extracted by these two
techniques, as well as through fitting to a dipole Ansatz is
demonstrated in \fig{fig:Final_Result_Iwasaki}. We show results for
the nEDM with the topological charge computed using either  cooling or the
gradient flow and the Iwasaki gauge action. As can be seen, results
extracted from fitting to a dipole Ansatz have very similar mean value to those
extracted  using continuum
derivative with the latter having smaller errors. The results
extracted from the elimination of the momentum in the plateau region tend to be  lower with larger
errors. This behavior resembles
  the results obtained for the  isovector magnetic form factor of the nucleon
at zero momentum \cite{Alexandrou:2014exa} where it was found that 
the results obtained using the elimination of the momentum in the plateau region method tend to be  larger (and closer to the experimental value) than those obtained using a dipole
fit .\par 

\begin{figure}[htb]
\vspace{-3cm}
\centerline{\hspace{0.0cm}\includegraphics[scale=0.6,angle=0]{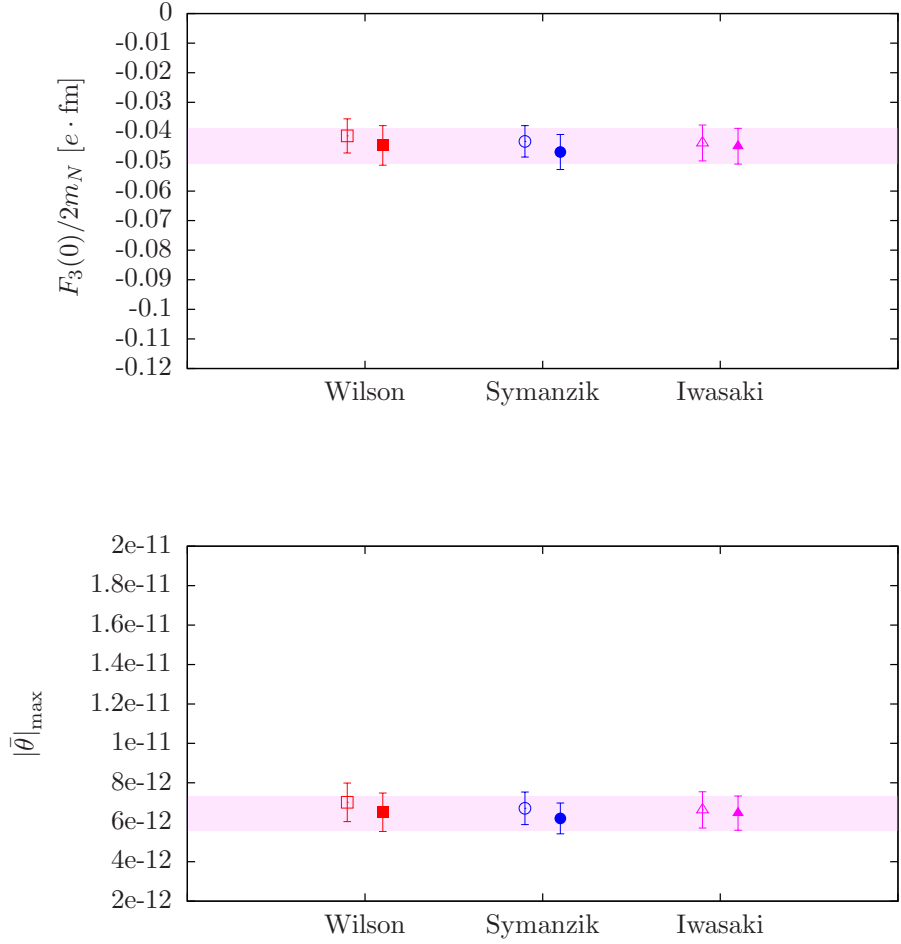}}
\vspace{-1.5cm}
\caption{\label{fig:Final_Result} 
Upper panel: Our results for $F_3(0)/(2m_N)$ in physical
units using different methods. Open (filled) symbols show results
using cooling (gradient flow) for the evaluation of the topological
charge. Red/blue/magenta points represent the smoothing actions
Wilson/Symanzik tree-level improved/Iwasaki). The presented results
have been obtained from the weighted averages of values for $F_3(0)/(2
m_N)$ extracted using: the dipole fit on $F_3(Q^2)$, the application of the derivative to the ratio technique,
and the elimination of the momentum in the plateau region technique. As a final result we report
the value of nEDM extracted when using ${\cal Q}$ from the gradient flow with Iwasaki action,
shown with the magenta error band.
Lower panel: The corresponding upper bounds in $\vert \bar{\theta} \vert$
extracted using the experimental result $\vert \vec{d}_N \vert = 2.9
\times 10^{-13} e \cdot {\rm fm}$ \cite{Helaine:2014ona,Baker:2006ts,Baker:2007df}.}
\end{figure}\FloatBarrier

An additional  new element of this work, is the computation of the
topological charge using both cooling and the gradient flow method. We
show that the two approaches are equivalent if the flow time and
number of cooling steps are tuned appropriately. This agreement is
demonstrated in  \fig{fig:Final_Result_Iwasaki} and \fig{fig:Final_Result} where we show results for the
nEDM with the topological charge computed using cooling or the
gradient flow method. Furthermore, results using different actions to
smooth the gauge links entering the computation of the topological
charge yield overall consistent results. The values appearing in
\fig{fig:Final_Result} have been obtained by taking the weighted
average among the data extracted using the dipole fit, the application of the derivative to the ratio technique and the elimination of the momentum in the plateau region, shown in \fig{fig:Final_Result_Iwasaki}. Given that the simulations
used the Iwasaki action, we present as our final value for the nEDM
the value extracted when the Iwasaki action is employed to define the topological charge.
As systematic error we take the difference between the mean values
obtained when cooling and the gradient flow are used to determine the
topological charge. Our final result is thus $F_3(0)/(2m_N) =
-0.045(6)(1)\, e\cdot {\rm fm}$ at a pion mass of $m_\pi=373$~MeV. As
already remarked in the introduction, this value is in agreement with
the value extracted using $N_f{=}2{+}1$ domain wall fermions at pion mass
of about 300~MeV~\cite{Shintani:2014zra}. Using this value of
$F_3(0)/(2m_N)$ and the experimental result $\vert \vec{d}_N \vert = 2.9
\times 10^{-13} e \cdot {\rm fm}$ as an upper bound we can extract the
maximum allowed value of ${\bar{\theta}}$ displayed in
Fig.~\ref{fig:Final_Result}. We find a maximum value of
${\bar{\theta}}=6.4(0.9)(0.2)\times 10^{-12}$.

Having investigated the nEDM using an ensemble simulated at $m_\pi=373$~MeV we are in the process of analyzing an ensemble with two degenerate light quarks with  mass fixed to the physical value, thus eliminating any systematic error that
may arise due to a heavier than physical pion mass.

\section*{Acknowledgments}
We would like to thank all members of ETMC for the most enjoyable
collaboration. Numerical calculations have used HPC resources from
John von Neumann-Institute for Computing on the JUQUEEN and JUROPA
systems at the research center in J\"ulich. This work was supported, in part, by a grant from the Swiss National Supercomputing Centre (CSCS) under project ID s540. Additional computational
resources were provided by the Cy-Tera machine at The Cyprus Institute
funded by the Cyprus Research Promotion Foundation (RPF), ${\rm
  NEAY\Pi O \Delta OMH}$/${\Sigma}$TPATH/0308/31. In addition, we
would like to thank H.~Panagopoulos, M.~Teper, B.~Lucini,
A.~Ramos and C. Michael for useful discussion on various
aspects of this project. K.O. is supported by the Bonn-Cologne Graduate
School (BCGS) of Physics and Astronomie. M.C. is supported by the
Cyprus RPF under the contract 
TECHNOLOGY/${\rm \Theta E\Pi I\Sigma}$/0311(BE)/16. K.H. acknowledges support by the Cyprus RPF
under contract ${\rm T \Pi E}$/${\rm \Pi \Lambda H P O}$/0311(BIE)/09.


\begin{thebibliography}{99}

\bibitem{Dar:2000tn}
 S.~Dar, {\it ``The Neutron EDM in the SM: A Review''}, hep-ph/0008248.

\bibitem{Pospelov:2005pr} 
M.~Pospelov and A.~Ritz, {\it ``Electric dipole moments as probes of new physics''}, Annals Phys.\  {\bf 318}, 119 (2005), [hep-ph/0504231].

\bibitem{Helaine:2014ona}
  V.~Helaine [nEDM Collaboration], {\it ``The neutron Electric Dipole Moment experiment at the Paul Scherrer Institute''},  EPJ Web Conf.\  {\bf 73} (2014) 07003.

\bibitem{Baker:2006ts} 
  C.~A.~Baker {\it et al.}, {\it ``An Improved experimental limit on the electric dipole moment of the neutron''}, Phys.\ Rev.\ Lett.\  {\bf 97}, 131801 (2006), [hep-ex/0602020].

\bibitem{Baker:2007df} 
  C.~A.~Baker {\it et al.},
  {\it ``Reply to comment on `An Improved experimental limit on the electric dipole moment of the neutron' ''}, Phys.\ Rev.\ Lett.\  {\bf 98}, 149102 (2007), [arXiv:0704.1354].

\bibitem{Baluni:1978rf} 
  V.~Baluni, {\it ``$CP$ Violating Effects in QCD''}, Phys.\ Rev.\ D {\bf 19}, 2227 (1979).

\bibitem{Crewther:1979pi} 
  R.~J.~Crewther, P.~Di Vecchia, G.~Veneziano and E.~Witten, {\it ``Chiral Estimate of the Electric Dipole Moment of the Neutron in Quantum Chromodynamics''}, Phys.\ Lett.\ B {\bf 88}, 123 (1979), [Phys.\ Lett.\ B {\bf 91}, 487 (1980)].

\bibitem{Shifman:1979if}
  M.~A.~Shifman, A.~I.~Vainshtein and V.~I.~Zakharov, {\it ``Can Confinement Ensure Natural $CP$ Invariance of Strong Interactions?''}, Nucl.\ Phys.\ B {\bf 166} (1980) 493.

\bibitem{Schnitzer:1983pb} 
  H.~J.~Schnitzer, {\it ``The Soft Pion Skyrmion Lagrangian and Strong $CP$ Violation''}, Phys.\ Lett.\ B {\bf 139}, 217 (1984).

\bibitem{Shabalin:1982sg} 
  E.~P.~Shabalin, {\it ``The Electric Dipole Moment Of The Neutron In A Gauge Theory''}, Sov.\ Phys.\ Usp.\  {\bf 26}, 297 (1983) [Usp.\ Fiz.\ Nauk {\bf 139}, 561 (1983)].

\bibitem{Musakhanov:1984qy} 
  M.~M.~Musakhanov and Z.~Z.~Israilov, {\it ``The Electric Dipole Moment Of The Neutron In The Chiral Bag Model''}, Phys.\ Lett.\ B {\bf 137}, 419 (1984).

\bibitem{Cea:1984qv} 
  P.~Cea and G.~Nardulli, {\it ``A Realistic Calculation of the Electric Dipole Moment of the Neutron Induced by Strong ${CP}$ Violation''}, Phys.\ Lett.\ B {\bf 144}, 115 (1984).

\bibitem{Morgan:1986yy}
  M.~A.~Morgan and G.~A.~Miller, {\it ``The Neutron Electric Dipole Moment in the Cloudy Bag Model''}, Phys.\ Lett.\ B {\bf 179} (1986) 379.

\bibitem{McGovern:1992bk} 
  J.~A.~McGovern and M.~C.~Birse, {\it ``The neutron electric dipole moment in chiral quark-meson models''}, Phys.~Rev.~{\bf D45}, 2437 (1992).

\bibitem{DiVecchia:1980gi} 
  P.~Di Vecchia, {\it ``The Dynamics of the Pseudoscalar Mesons at Arbitrary $\Theta$ in Large $N$ Quantum Chromodynamics''}, Acta Phys.\ Austriaca Suppl.\  {\bf 22}, 341 (1980).

\bibitem{Pich:1991fq}
 A.~Pich and E.~de Rafael, {\it ``Strong CP violation in an effective chiral Lagrangian approach''}, Nucl.\ Phys.\ B {\bf 367} 313 (1991).

\bibitem{Borasoy:2000pq} 
 B.~Borasoy, {\it ``The Electric dipole moment of the neutron in chiral perturbation theory''}, Phys.\ Rev.\ D {\bf 61}, 114017 (2000), [arXiv:hep-ph/0004011].

\bibitem{Hockings:2005cn} 
 W.~H.~Hockings and U.~van Kolck, {\it ``The Electric dipole form factor of the nucleon''}, Phys.\ Lett.\ B {\bf 605}, 273 (2005), [arXiv:nucl-th/0508012].

\bibitem{Narison:2008jp} 
 S.~Narison, {\it ``A Fresh Look into the Neutron EDM and Magnetic Susceptibility''}, Phys.\ Lett.\ B {\bf 666}, 455 (2008), [arXiv:0806.2618].

\bibitem{Ottnad:2009jw}
 K.~Ottnad, B.~Kubis, U.-G.~Meissner and F.-K.~Guo, {\it``New insights into the neutron electric dipole moment''}, Phys.\ Lett.\ B {\bf 687} 42 (2010), [arXiv:0911.3981].

\bibitem{deVries:2010ah} 
 J.~de Vries, R.~G.~E.~Timmermans, E.~Mereghetti and U.~van Kolck, {\it ``The Nucleon Electric Dipole Form Factor From Dimension-Six Time-Reversal Violation''}, Phys.\ Lett.\ B {\bf 695}, 268 (2011), [arXiv:1006.2304].

\bibitem{Mereghetti:2010kp} 
 E.~Mereghetti, J.~de Vries, W.~H.~Hockings, C.~M.~Maekawa and U.~van Kolck, {\it``The Electric Dipole Form Factor of the Nucleon in Chiral Perturbation Theory to Sub-leading Order''}, Phys.\ Lett.\ B {\bf 696}, 97 (2011), [arXiv:1010.4078].

\bibitem{deVries:2012ab} 
 J.~de Vries, E.~Mereghetti, R.~G.~E.~Timmermans and U.~van Kolck, {\it``The Effective Chiral Lagrangian From Dimension-Six Parity and Time-Reversal Violation''}, Annals Phys.\  {\bf 338}, 50 (2013), [arXiv:1212.0990].

\bibitem{Guo:2012vf} 
 F.~K.~Guo and U.~G.~Meissner, {\it ``Baryon electric dipole moments from strong CP violation''}, JHEP {\bf 1212}, 097 (2012), [arXiv:1210.5887].

\bibitem{Akan:2014yha} 
 T.~Akan, F.~K.~Guo and U.~G.~Meissner, {\it``Finite-volume corrections to the CP-odd nucleon matrix elements of the electromagnetic current from the QCD vacuum angle''}, Phys.\ Lett.\ B {\bf 736}, 163 (2014), [arXiv:1406.2882].

\bibitem{Peccei:1977hh} 
  R.~D.~Peccei and H.~R.~Quinn, {\it ``CP Conservation in the Presence of Instantons''}, Phys.\ Rev.\ Lett.\  {\bf 38}, 1440 (1977).

\bibitem{Peccei:1988ci} 
  R.~D.~Peccei, {\it ``The Strong {CP} Problem''}, Adv.\ Ser.\ Direct.\ High Energy Phys.\  {\bf 3}, 503 (1989).

\bibitem{Jarlskog:1988}
S. M. Barr and W. J. Marciano in {\it ``CP Violation''}, ed. C Jarlskog (World Scientific, Singapore, 1988).

\bibitem{Alexandrou:2014exa}
C.~Alexandrou {\it et al.}, {\it ``Extraction of the isovector magnetic form factor of the nucleon at zero momentum''}, PoS LATTICE2014 (2014) 075, [arXiv:1410.8818].

\bibitem{Luscher:2010iy} 
  M.~L\"uscher, {\it ``Properties and uses of the Wilson flow in lattice QCD''}, JHEP {\bf 1008}, 071 (2010) [JHEP {\bf 1403}, 092 (2014)], [arXiv:1006.4518].

\bibitem{Alexandrou:2015yba}
C.~Alexandrou, A.~Athenodorou and K.~Jansen, { {\it ``Topological charge using cooling and the gradient flow''}, {Phys.\ Rev.\ D {\bf 92} (2015) 12,  125014}, [arXiv:1509.04259].}

\bibitem{Shintani:2014zra} 
  E.~Shintani, T.~Blum, A.~Soni and T.~Izubuchi,
  {\it ``Neutron and proton EDM with $N_{f} = 2 + 1$ domain-wall fermion''}, PoS LATTICE {\bf 2013}, 298 (2014).

\bibitem{Shintani:2008nt} 
  E.~Shintani, S.~Aoki and Y.~Kuramashi, {\it ``Full QCD calculation of neutron electric dipole moment with the external electric field method''}, Phys.\ Rev.\ D {\bf 78}, 014503 (2008), [arXiv:0803.0797].

\bibitem{Guo:2015tla} 
  F.-K.~Guo {\it et al.}, {\it ``The electric dipole moment of the neutron from 2+1 flavor lattice QCD''}, Phys.\ Rev.\ Lett.\  {\bf 115}, no. 6, 062001 (2015), [arXiv:1502.02295].

\bibitem{Shindler:2015aqa} 
A.~Shindler, T.~Luu and J.~de Vries,
{\it ``The nucleon electric dipole moment with the gradient flow: the $\theta$-term contribution''}, [arXiv:1507.02343].

\bibitem{Iwasaki:1996sn}
  Y.~Iwasaki, K.~Kanaya, T.~Kaneko and T.~Yoshie, {\it ``Scaling in SU(3) pure gauge theory with a renormalization group improved action,''}x
  Phys.\ Rev.\ D {\bf 56} (1997) 151 [hep-lat/9610023].

\bibitem{Frezzotti:2000nk}
 R.~Frezzotti {\it et al.} [Alpha Collaboration], {\it ``Lattice QCD with a chirally twisted mass term,''}
 JHEP {\bf 0108} (2001) 058
 [hep-lat/0101001].

\bibitem{Frezzotti:2003ni}
R.~Frezzotti and G.~C.~Rossi, {\it ``Chirally improving Wilson fermions - I. O(a) improvement''}, JHEP {\bf 0408} (2004) 007, [arXiv:hep-lat/0306014].

\bibitem{Frezzotti:2005gi} 
  {R.~Frezzotti, G.~Martinelli, M.~Papinutto and G.~C.~Rossi, {\it ``Reducing cutoff effects in maximally twisted lattice QCD close to the chiral limit,''} JHEP {\bf 0604}, 038 (2006) [hep-lat/0503034].}

\bibitem{Frezzotti:2003xj}
 R.~Frezzotti and G.~C.~Rossi, {\it ``Twisted mass lattice QCD with mass nondegenerate quarks''},
 Nucl.\ Phys.\ Proc.\ Suppl.\  {\bf 128} (2004) 193
 [hep-lat/0311008].

\bibitem{Baron:2010bv} 
  R.~Baron {\it et al.}, {\it ``Light hadrons from lattice QCD with light (u,d), strange and charm dynamical quarks''}, JHEP {\bf 1006}, 111 (2010), [arXiv:1004.5284].

\bibitem{Athenodorou:2011zp} 
  A.~Athenodorou and R.~Sommer, {\it ``One-loop lattice artifacts of a dynamical charm quark''} Phys.\ Lett.\ B {\bf 705}, 393 (2011), [arXiv:1109.2303].

\bibitem{Bruno:2014ufa} 
  M.~Bruno {\it et al.} [ALPHA Collaboration], {\it ``Effects of Heavy Sea Quarks at Low Energies''} Phys.\ Rev.\ Lett.\  {\bf 114}, no. 10, 102001 (2015), [arXiv:1410.8374].

\bibitem{Alexandrou:2014sha} C. Alexandrou {\it et al.}, {\it ``Hyperon and charmed baryon masses and nucleon excited states from lattice QCD''}, arXiv:1410.4553.

\bibitem{Baron:2011sf} 
 R.~Baron {\it et al.}  [ETM Collaboration], {\it``Light hadrons from $N_f{=}2{+}1{+}1$ dynamical twisted mass fermions''}, PoS LATTICE {\bf 2010}, 123 (2010), [arXiv:1101.0518].

\bibitem{Aoki:1989rx} 
  S.~Aoki and A.~Gocksch, {\it ``The Neutron Electric Dipole Moment in Lattice {QCD}''},  Phys.\ Rev.\ Lett.\  {\bf 63}, 1125 (1989) [Phys.\ Rev.\ Lett.\  {\bf 65}, 1172 (1990)].

\bibitem{Guadagnoli:2002nm} 
  D.~Guadagnoli, V.~Lubicz, G.~Martinelli and S.~Simula, {\it ``Neutron electric dipole moment on the lattice: A Theoretical reappraisal''} JHEP {\bf 0304}, 019 (2003), [hep-lat/0210044].

\bibitem{Facciolia:2004jz} 
  P.~Faccioli, D.~Guadagnoli and S.~Simula, {\it ``The Neutron electric dipole moment in the instanton vacuum: Quenched versus unquenched simulations''},  Phys.\ Rev.\ D {\bf 70}, 074017 (2004), [hep-ph/0406336].

\bibitem{Shintani:2005xg} 
  E.~Shintani {\it et al.}, {\it ``Neutron electric dipole moment from lattice QCD''},  Phys.\ Rev.\ D {\bf 72}, 014504 (2005), [hep-lat/0505022].

\bibitem{Shintani:2006xr} 
  E.~Shintani {\it et al.}, {\it ``Neutron electric dipole moment with external electric field method in lattice QCD''} Phys.\ Rev.\ D {\bf 75}, 034507 (2007), [hep-lat/0611032].

\bibitem{Alles:2005ys} 
  B.~Alles, M.~D'Elia and A.~Di Giacomo, {\it ``An Upper limit to the electric dipole moment of the neutron from lattice QCD''}, Nucl.\ Phys.\ Proc.\ Suppl.\  {\bf 164}, 256 (2007), [hep-lat/0510067].

\bibitem{Aoki:2008gv} 
  S.~Aoki, R.~Horsley, T.~Izubuchi, Y.~Nakamura, D.~Pleiter, P.~E.~L.~Rakow, G.~Schierholz and J.~Zanotti, {\it ``The Electric dipole moment of the nucleon from simulations at imaginary vacuum angle theta''} arXiv:0808.1428].

\bibitem{Stathopoulos:2007zi} 
  A.~Stathopoulos and K.~Orginos, {\it ``Computing and deflating eigenvalues while solving multiple right hand side linear systems in quantum chromodynamics''}, SIAM J.\ Sci.\ Comput.\  {\bf 32}, 439 (2010), [arXiv:0707.0131].

\bibitem{Stathopoulos:2009zz} 
A.~Stathopoulos, A.~M.~Abdel-Rehim and K.~Orginos, {\it ``Deflation for inversion with multiple right-hand sides in QCD''}, J.\ Phys.\ Conf.\ Ser.\  {\bf 180}, 012073 (2009).

\bibitem{Abdel-Rehim:2013wlz}
A. Abdel-Rehim {\it et al.}, {\it ``Disconnected quark loop contributions to nucleon observables in lattice QCD''}, Phys.Rev. {\bf D89} (2014) 034501, [arXiv:1310.6339].

\bibitem{Alexandrou:1992ti}
C.~Alexandrou {\it et al.}, {\it ``The Static Approximation of
  Heavy-Light Quark-Systems - A Systematic Lattice Study''}, Nucl. Phys. {\bf B414} (1994) 815-855, [arXiv:hep-lat/9211042].

\bibitem{Gusken:1989qx} 
S.~Gusken, {\it ``A Study of smearing techniques for hadron correlation functions''} Nucl.\ Phys.\ Proc.\ Suppl.\  {\bf 17}, 361 (1990).

\bibitem{Alexandrou:2008tn}
C.~Alexandrou {\it et al.}, {\it ``Light baryon masses with dynamical twisted mass fermions''}, Phys. Rev. {\bf D78} (2008) 014509, [arXiv:0803.3190].

\bibitem{Constantinou:2009tr} 
M.~Constantinou, V.~Lubicz, H.~Panagopoulos and F.~Stylianou, {\it ``${\cal O}(a^2)$ corrections to the one-loop propagator and bilinears of clover fermions with Symanzik improved gluons''}, JHEP {\bf 0910}, 064 (2009), [arXiv:0907.0381].

\bibitem{Alexandrou:2013joa} 
C.~Alexandrou, M.~Constantinou, S.~Dinter, V.~Drach, K.~Jansen, C.~Kallidonis and G.~Koutsou, {\it ``Nucleon form factors and moments of generalized parton distributions using $N_f=2+1+1$ twisted mass fermions''}, Phys.\ Rev.\ D {\bf 88}, no. 1, 014509 (2013), [arXiv:1303.5979].

\bibitem{Alexandrou:2015dc}
C.~Alexandrou {\it et al.}, {\it ``Direct computation of the nucleon magnetic moment''}, in preparation.

\bibitem{Luscher:2009eq} 
M.~Luscher, {\it ``Trivializing maps, the Wilson flow and the HMC algorithm''},  Commun.\ Math.\ Phys.\  {\bf 293}, 899 (2010), [arXiv:0907.5491].

\bibitem{Bonati:2014tqa} 
C.~Bonati and M.~D'Elia, {\it ``Comparison of the gradient flow with cooling in $SU(3)$ pure gauge theory''}, Phys.\ Rev.\ D {\bf 89}, no. 10, 105005 (2014), [arXiv:1401.2441].

\bibitem{Anderson:1954}
T.W. Anderson and D.A. Darling,  {\it "A Test of Goodness-of-Fit"}, Journal of the American Statistical Association {\bf 49} 765–769 (1954).

\bibitem{Bernecker:2011gh} 
D.~Bernecker and H.~B.~Meyer,
{\it ``Vector Correlators in Lattice QCD: Methods and applications''},
Eur.\ Phys.\ J.\ A {\bf 47}, 148 (2011), [arXiv:1107.4388].

\bibitem{Feng:2013xsa} 
X.~Feng et al., {\it ``Computing the hadronic vacuum polarization function by analytic continuation''}, Phys.\ Rev.\ D {\bf 88}, 034505 (2013), [arXiv:1305.5878].

\end{thebibliography}
\end{document}